\documentclass{aastex63}

\usepackage{graphicx}
\usepackage{txfonts}

\received{June 1, 2019}
\revised{January 10, 2019}
\accepted{\today}
\submitjournal{ApJ}

\shorttitle{Diagnostic of eruptive active regions}
\shortauthors{Pagano et al.}

\begin{document}

\title{A Prospective New Diagnostic Technique for Distinguishing Eruptive and Non-Eruptive Active Regions}

\correspondingauthor{Paolo Pagano}
\email{pp25@st-andrews.ac.uk}

\author{Paolo Pagano}
\affil{School of Mathematics and Statistics,
University of St Andrews, North Haugh
St Andrews, KY16 9SS, Scotland}

\author{Duncan H. Mackay}
\affil{School of Mathematics and Statistics,
University of St Andrews, North Haugh
St Andrews, KY16 9SS, Scotland}

\author{Stephanie L. Yardley}
\affil{School of Mathematics and Statistics,
University of St Andrews, North Haugh
St Andrews, KY16 9SS, Scotland}



\begin{abstract}

Active regions are the source of the majority of magnetic flux rope ejections that become Coronal Mass Ejections (CMEs). To identify in advance which active regions will produce an ejection is key for both space weather prediction tools and future science missions such as Solar Orbiter. The aim of this study is to develop a new technique to identify which active regions are more likely to generate magnetic flux rope ejections. The new technique will aim to: (i) produce timely space weather warnings and (ii) open the way to a qualified selection of observational targets for space-borne instruments. We use a data-driven Non-linear Force-Free Field (NLFFF) model to describe the 3D evolution of the magnetic field of a set of active regions. We determine a metric to distinguish eruptive from non-eruptive active regions based on the Lorentz force. Furthermore, using a subset of the observed magnetograms, we run a series of simulations to test whether the time evolution of the metric can be predicted. The identified metric successfully differentiates active regions observed to produce eruptions from the non-eruptive ones in our data sample. A meaningful prediction of the metric can be made between 6 to 16 hours in advance. This initial study presents an interesting first step in the prediction of CME onset using only LOS magnetogram observations combined with NLFFF modelling. Future studies will address how to generalise the model such that it can be used in a more operational sense and for a variety of simulation approaches.

\end{abstract}



\section{Introduction}
\label{introduction}

Identifying potential sources of solar eruptions has recently become
one of the main goals for solar physics. The accurate prediction of the
source region and, ideally, the onset time of eruptions is essential for scientific missions, such as Solar Orbiter and
for the development of a new generation of space weather forecasting models.
The consequences of space weather have been extensively studied and we refer to \citet{Schrijver2015} for a detailed discussion.
It is generally believed that the prediction of the onset of coronal mass ejections (CMEs) is key to mitigating the consequences of Space Weather. Advance knowledge of
the lift off time of CMEs (and consequently the arrival time at Earth)
is required to react accordingly to space weather threats.
Most governments have included space weather into their national risks analysis 
(e.g. UK \textit{National Risk Register of Civil Emergencies})
and counter measures are being taken worldwide to mitigate its effects
\textit{(e.g. The Space weather preparedness strategy}).
For missions such as Solar Orbiter,
the identification of the eruption source region is important as relevant observational targets need to be identified days in advance.
At present, warnings are issued when observational signatures
of eruptions are detected. However, earlier identification would 
lead to warnings being issued a few hours prior to eruptions.
This is a minimum requirement
to (i) produce meaningful alerts, (ii)  
run basic models to infer the properties and trajectories of the resulting CMEs and (iii)
re-point telescopes. While a few hours warning is the minimum requirement, in the long term, predictions of CME onset a few days in advance is desirable.

Magnetic flux ropes are twisted magnetic structures found in the solar corona that connect opposite polarities and tend to lie along polarity inversion lines \citep[PILs][]{Cheng2010}.
Theoretical models suggest that either a weakly twisted magnetic flux rope \citep{Isenberg1993, Amari2000} or a highly sheared arcade \citep{Ouyang2015} are a necessary ingredient to form CMEs,
as these are the only structures that can store the necessary amount of free magnetic energy that is then abruptly released.
However, it is not always possible to determine the pre-eruptive magnetic configuration of CMEs.
In some cases observational evidence suggests that a magnetic flux rope is present prior to eruption \citep{Chen2011,Yan2017, Song2014, HowardDeForest2014},
while in others it is too difficult to reach a conclusion from the observations.
Active regions, due to their intense and complex magnetic fields, are the preferred locations for the formation and ejection of magnetic flux ropes.
It is important to analyse the 3D magnetic field evolution and the stability of these pre-eruptive structures to identify the onset of eruption.

As it is currently very difficult to measure the magnetic field in the solar corona,
to understand the 3D coronal
magnetic configuration, extrapolations from normal component magnetograms at the photosphere are required
\citep[for reviews][]{MackayYeates2012,Wiegelmann2017}.
This provides a representation of the coronal field at a single instant in time.
In simple terms, these models reconstruct a magnetohydrostatic equilibrium
that satisfy prescribed boundary conditions where the magnetic field can have different degrees of complexity (from potential to NLFFF).
A subset of these models uses the magnetofrictional relaxation technique \citep{Yang1986},
where a continuous series of NLFFF equilibria are produced.
The data-driven NLFFF model of \citet{Mackay2011} uses a time series of normal component magnetograms as the lower boundary conditions
to produce a quasi-static evolution of the coronal magnetic field through a series of near equilibrium states.
This approach has been shown to be accurate in describing the non-potential field above active regions.
It has been successful in reproducing sigmoids and the formation of magnetic flux ropes
\citep{Gibb2014, Yardley2018a, Yardley2019}
along with the generation of conditions required for the onset of magnetic flux rope ejections \citep{MackayVanBallegooijen2006A,Yeates2010,Pagano2013a,Pagano2013b,Pagano2014,Rodkin2017}.

In this paper, we use the model of \cite{Mackay2011} to develop a new technique aimed at identifying which active regions are most likely to produce an eruption.
We apply this model to a set of active regions that have been previously studied in detail \citep{Rodkin2017,Yardley2018a,Yardley2018b,James2018,Yardley2019}.
Some of the active regions resulted in observed eruptions, while others did not.
We first analyse the 3D magnetic field configuration of the active regions produced by the magnetofrictional model
to identify a metric that discriminates the eruptive and non-eruptive active regions.
Once this metric is identified, we focus on the predictive capabilities of this approach.
Magnetofrictional simulations are run where
the photospheric magnetic field evolution is projected forward in time
without further input from magnetograms.
The method is continued forward for up to 32 hours to see whether
the eruptive or non-eruptive state of the active regions can be predicted correctly.
Two techniques are considered (i) simply using the present evolution to project the future evolution and, (ii) adding a component of noise in the projection of the magnetograms to test its robustness.
This allows us to consider how the projected evolution affects the metric.

The structure of the paper is as follows: in Sec.\ref{model} we give more details on the magnetofrictional model and the active regions under study.
In Sec.\ref{criteria} we discuss the parameters that differentiate active regions with observed eruptions from those without.
Next  in Sec.\ref{prediction} we show how the use of projected magnetograms affect this application
and we draw some conclusions in Sec.\ref{conclusions}.

\section{Model and simulations}
\label{model}

The work presented here
is based on the magnetofrictional simulation approach of \cite{Mackay2011}
where a continuous time series of 3D NLFFF configurations
are derived from a corresponding time series of magnetogram measurements.
We apply this model to eight active regions, where
in five of these regions eruptions were observed,
while for the remaining three regions no eruptions occurred.

\subsection{Model}
\label{hexa}
The magnetofrictional simulation describes the magnetic field evolution in a cartesian 3D domain by considering the simultaneous stressing and relaxation of the coronal magnetic field.
The stressing of the field is due to the evolution of the magnetic flux distribution at the lower photospheric boundary determined from a time series of magnetogram observations.
The relaxation occurs from specifying the velocity
to be proportional to the Lorentz force in the 3D domain.
Full details of the NLFFF model can be found in \citet{Mackay2011}.
In this model, the 3D domain is a cartesian box where the solar surface is placed at the lower z-boundary.
The horizontal directions, $x$ and $y$, extend over a sufficient region of the solar surface to fully contain the active region.
In this study, the time series of NLFFF configurations is constructed assuming
closed boundaries at the four sides of the 3D box (no normal magnetic field),
while the magnetic field can have a normal component across the top boundary.
The bottom boundary, which represents the solar surface, is forced to have evolving and balanced magnetic flux.

The model uses a zero-$\beta$ approximation where this
provides an accurate representation of the evolution of the coronal magnetic field over times scales longer than the Alf\'ven crossing time.
The initial coronal magnetic field for each active region is assumed to be a  potential
field and at later times the evolution of the magnetic field at the lower boundary (derived from observed line-of-sight magnetograms)
leads to the injection and build-up of electric currents in the corona.
Thus, the coronal magnetic field evolves to a new NLFFF configuration.
It is the evolution of the magnetic flux at the lower boundary that is 
key to the build up of magnetic forces in the domain during this quasi-static evolution.
The relaxation of the magnetic configuration is tuned to match the relaxation times in the solar corona.

Occasionally, during the quasi-static evolution, the model cannot converge to a new 
NLFFF equilibrium due to the build-up of large flux ropes. This usually occurs in conjunction 
with the lift off of a magnetic flux rope in the model,
where magnetic reconnection below the flux rope leads to a strong outward magnetic tension \citep{MackayVanBallegooijen2006B}.
At this point the NLFFF model is no longer appropriate and full MHD is required to describe the correct dynamics \citep{Pagano2013a}.

\subsection{Observed Active Regions and Eruptions}
\label{event}

In order to develop a technique to identify active regions in which 
magnetic flux rope ejections occur,
we consider a number of active regions that have previously been analysed in detail.
In five of these active regions
observable signatures of eruptions have been clearly identified
and the other three show no such signatures.
\citet{Yardley2018b} provides an overview of what observable signatures
can be interpreted as the occurrence of an eruption in an active region.
Table \ref{activeregions} shows the main properties of the active regions
selected for this study
and we refer to them as eruptive or non-eruptive active regions as
appropriate \citep{Rodkin2017,Yardley2018a,Yardley2018b,James2018,Yardley2019}.
To identify eruptions we focus mostly on dynamic signatures found within coronal 
images that indicate a rapid plasma displacement or ejection. These signatures include
coronal dimmings, filament eruptions, the disappearance of coronal loops, or post 
flare magnetic field rearrangement\citep{Yardley2018b}. While CMEs can be linked to solar flares, 
both phenomena can occur without the other \citep{Gopalswamy2004}. Due to this we do not 
use GOES data which is more related to burst heating or energetic particles compared to
a large scale displacement or re-arrangement in the coronal field.
For the present study, we have favoured active regions isolated from large 
concentrations of magnetic flux in order to simplify the analysis.
Each of the active regions is observed to undergo qualitatively different evolution over the time period studied. 
Some of them (e.g. AR11504 and AR11561) show clear indications of magnetic flux emergence, while others do not.

\begin{table}
\centering                          
\begin{tabular}{c c c c c c}        
\hline\hline                 
 Active region & Observation Start & Observation End & Eruption time and signatures & Publication \\    
\hline                        
AR11561 & 2012.08.29 19:12:05 S18 E34 & 2012.09.02 01:36:04 S20 W12 & 2012.09.01 23:37 CD, EL, FA, FR & Y18b, Y19 \\  
AR11680 & 2013.02.24 14:23:55 S25 E52 & 2013.03.03 19:11:56 S24 W38 & 2013.03.03 17:27 CD, FA, FE, FR & Y18b, Y19 \\
AR11437 & 2012.03.16 12:47:57 S29 E33 & 2012.03.21 01:35:58 S29 W21 & 2012.03.20 14:46 CD, EL, FA & Y18a, Y18b, Y19 \\
AR11261 & 2011.07.31 05:00:41 N10 E18 & 2011.08.02 06:00:41 N10 W12 & 2011.08.02 05:54 EL, FA, FR & R17 \\
AR11504 & 2012.06.11 00:00:08 S18 E55 & 2012.06.14 22:24:08 S18 E13  & 2012.06.14 13:52 CD, EL, FA, FR & J18 \\
AR11480 & 2012.05.09 11:12:05 S14 E26 & 2012.05.14 00:00:05 S14 W36 & none & Y18b, Y19 \\  
AR11813 & 2013.08.06 01:36:07 S19 E22 & 2013.08.12 00:00:07 S17 W63 & none & Y18b, Y19 \\
AR12455 & 2015.11.13 04:47:55 N14 E61 & 2015.11.18 23:59:54 N13 W17 & none & Y18b, Y19 \\

\hline                                   
\end{tabular}
\caption{Properties of the active regions analysed in this study. The magnetogram cadence is 96 minutes for all, except AR11261 where it is 60 minutes.
Eruption signatures legend is as follows: CD (Coronal Dimmings), EL (Rapid disappearance of coronal loops in EUV), FA (Flare Arcade), FE (Filament Eruption), FR (Flare Ribbons).
Publications legend is: R17 \citep{Rodkin2017}, Y18a \citep{Yardley2018a}, Y18b \citep{Yardley2018b}, J18 \citep{James2018}, Y19 \citep{Yardley2019}.}
\label{activeregions}
\end{table}

For each of the active regions, we simulate the evolution of the 3D coronal field over the time period given in Table \ref{activeregions}, using the modelling approach discussed in \citet{Mackay2011}
and the follow up works of
\citet{Mackay2011}, \citet{Gibb2014}, \citet{Rodkin2017}, \citet{Yardley2018a}, and \citet{Yardley2019}.
For the present study,
we do not analyse the evolution of the coronal field in detail,
but we focus on defining a metric based on the evolution of the magnetic field configuration and the vertical component of the Lorentz force, $LF_z$,  to identify eruptive active regions.

\begin{figure}
\centering
\includegraphics[scale=0.25]{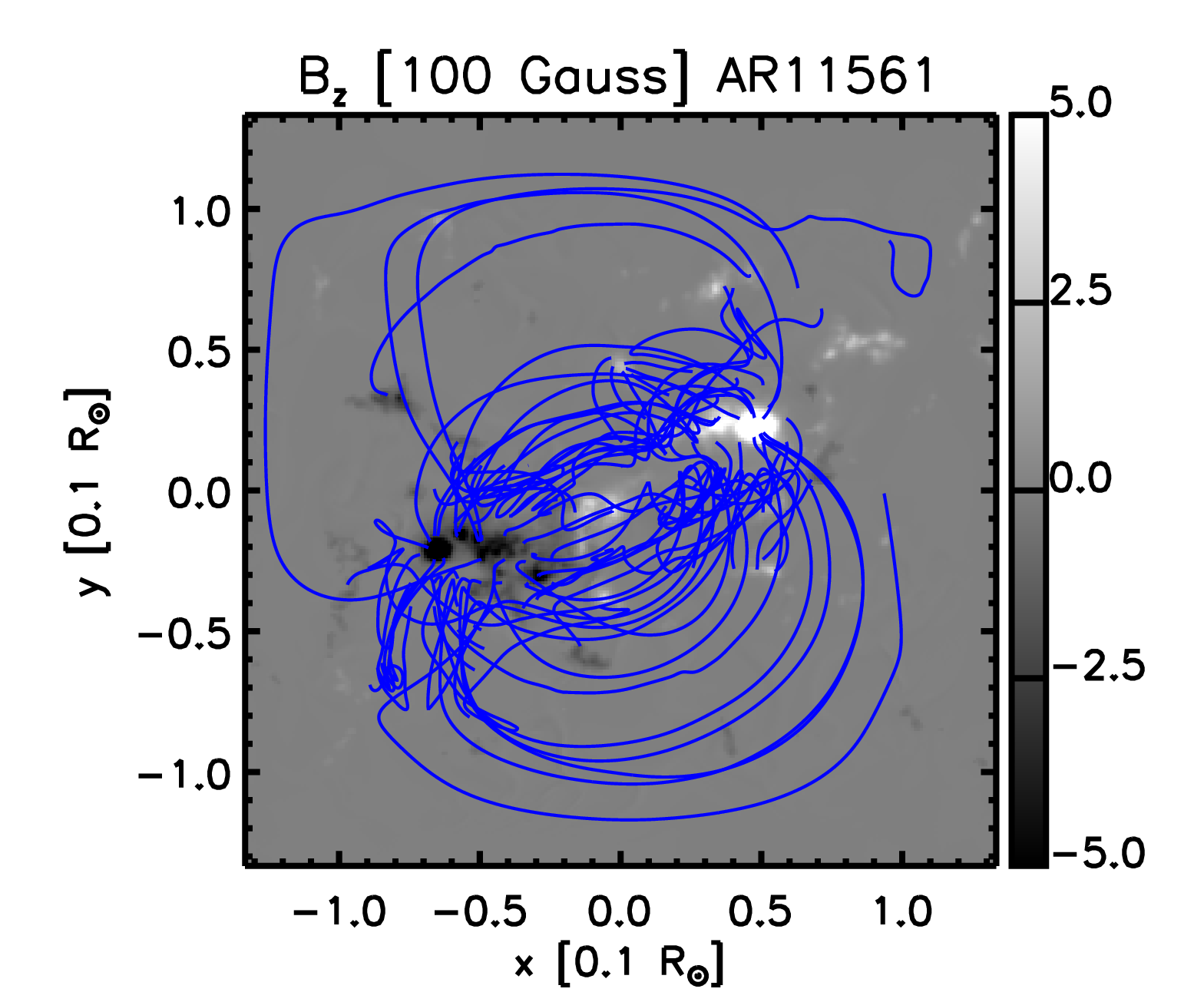}
\includegraphics[scale=0.25]{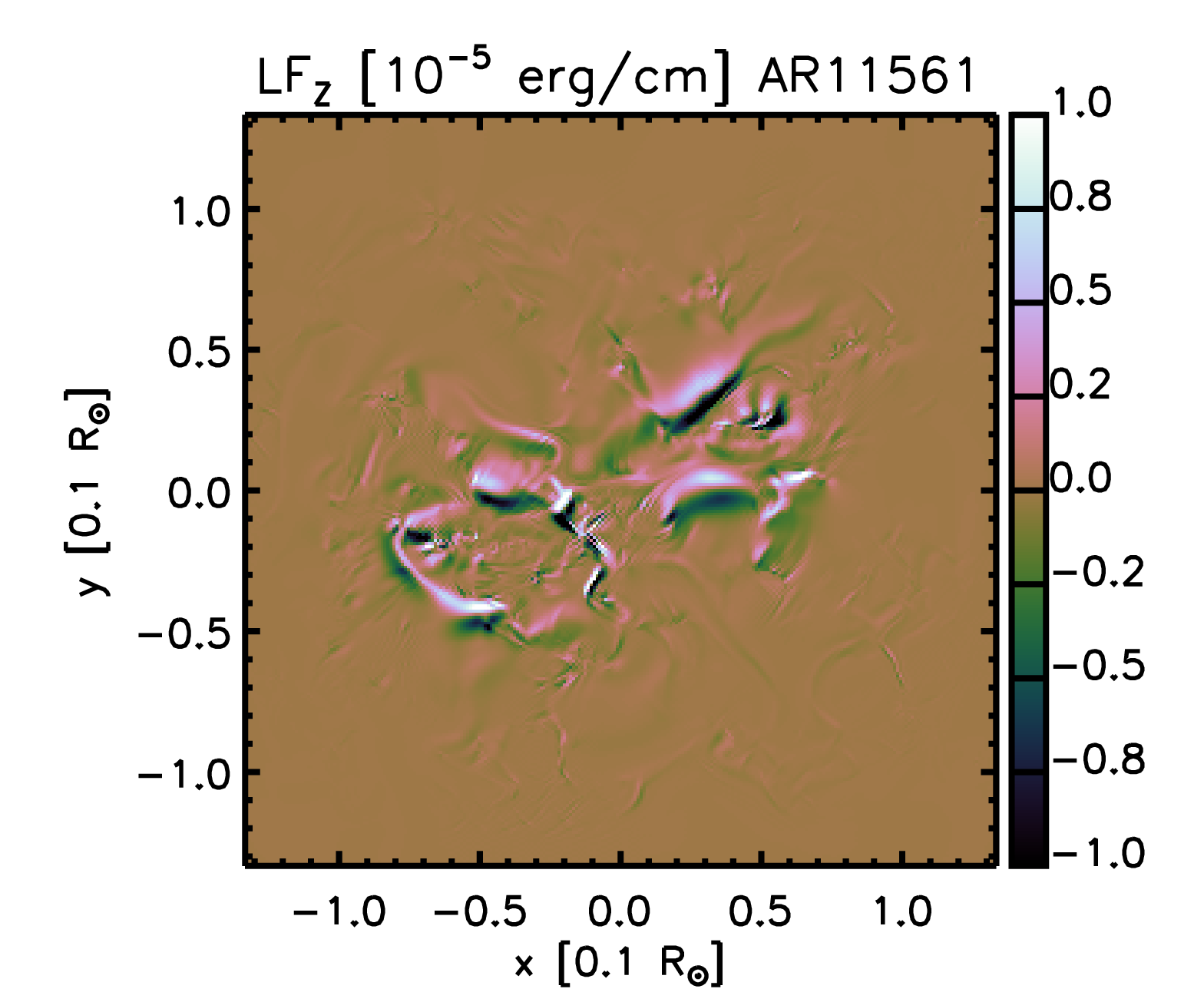}

\caption{Left-hand side: map of the vertical component of the magnetic filed ($B_z$) at the lower boundary of the magnetofrictional simulation at the time of an observed eruption for AR11561
with magnetic field lines from the model overplotted.
Right-hand side: corresponding map of the vertical components of the Lorentz force ($LF_z$).}
\label{bzeruptive}
\end{figure}

Fig.\ref{bzeruptive} presents a typical example of the output of our model.
The left-hand side panel shows the magnetic field distribution at the solar surface along with magnetic field lines from the model overplotted.
The right-hand side shows the associated vertical component of the Lorentz force at the same surface.
Initial studies show that simply using these 2D maps at the 
lower boundary cannot distinguish eruptive from non-eruptive active regions.

\section{Flux rope ejection metric}
\label{criteria}

In this section, we analyse the 3D magnetic field configuration of the active regions
simulated using the magnetofrictional model to identify
a metric that distinguishes eruptive from non-eruptive active regions.
We focus on the magnetic field configuration and the Lorentz force,
as they are ultimately responsible for triggering
the onset of magnetic flux rope ejections.

In order to discriminate eruptive and non-eruptive active regions,
we introduce an eruption metric $\zeta$
as the product of three different properties of the magnetic field configuration.
\begin{equation}
\zeta=\omega\mu\sigma,
\label{zeta}
\end{equation}
where $\omega$ is a proxy for the formation of a magnetic flux rope,
$\mu$ is connected to the vertical component of the Lorentz force,
and $\sigma$ represents the heterogeneity of the Lorentz force.
In the following subsections, we describe in details
the computation and physical interpretation of the functions
$\omega$, $\mu$, and $\sigma$.

\subsection{Flux ropes occurrence - $\omega$}
\label{omegaderivation}
We first adopt a proxy for the formation of magnetic flux ropes
previously applied in \citet{Rodkin2017}
and \citet{Pagano2018}. This approach allows us to track magnetic flux ropes using the function:
\begin{equation}
\Omega=\sqrt{\Omega_x^2+\Omega_y^2+\Omega_z^2}
\end{equation}
where
\begin{equation}
\Omega_x=\frac{\left|B\times\nabla B_x\right|}{\left|\nabla B_x\right|},
\end{equation}
\begin{equation}
\Omega_y=\frac{\left|B\times\nabla B_y\right|}{\left|\nabla B_y\right|},
\end{equation}
\begin{equation}
\Omega_z=\frac{\left|B\times\nabla B_z\right|}{\left|\nabla B_z\right|}.
\end{equation}
The function $\Omega$ depends on the twist of the magnetic field and its strength.
For example, $\Omega_z$ peaks at PILs where gradients of $B_z$ are perpendicular to the direction of $\vec{B}$.
Such a function is useful for identifying where flux ropes have formed or are about to form.
To produce time dependent 2D maps that represent the location on the surface where flux ropes exist in the corona we consider $\omega^*\left(x,y,t\right)$,
the integral of $\Omega$ along the z direction,
\begin{equation}
\omega^*\left(x,y,t\right)=\int_{z=0}^{z=z_{max}} \Omega\left(x,y,z,t\right) dz,
\label{eq:omega}
\end{equation}
where $z=0$ is the lower boundary of the computational box and $z=z_{max}$ is the upper boundary.
\begin{figure}
\centering
\includegraphics[scale=0.25]{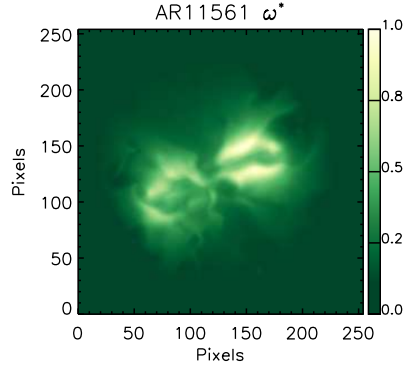}
\caption{Map of the function $\omega^*\left(x,y,t\right)$ (Eq.\ref{eq:omega}) for AR11561 at the time
when an eruption is observed.
The value of $\omega^*\left(x,y,t\right)$ is normalised with respect to the maximum of the function at this time.}
\label{omegaAR11561}
\end{figure}
Fig.\ref{omegaAR11561} shows a map of the function $\omega^*\left(x,y,t\right)$ normalised to its maximum value for the eruptive
active region AR11561. Typically, we find that $\omega^*\left(x,y,t\right)$
has significantly larger values
at a few locations located across the active region,
but it also has non-zero values over a large portion of the active region.

Finally, we need to obtain a normalised distribution, $\omega\left(x,y,t\right)$, of the quantity $\omega^*\left(x,y,t\right)$
that can be used for the derivation of the metric $\zeta$.
Thus, we exclude a frame of 16 pixels near the x and y boundaries to avoid boundary effects,
and we renormalise the functions $\omega^*$
to be between the values of 0 and 1.
\begin{equation}
\omega\left(x,y,t\right)=\frac{\omega^*\left(x,y,t\right)-min(\omega^*\left(x,y,t'\le t\right))}{max(\omega^*\left(x,y,t'\le t\right))-min(\omega^*\left(x,y,t'\le t\right))}
\label{orenorm}
\end{equation}
It should be noted that this normalisation is time dependent, as at time $t$
the function is normalised with respect to the maximum and minimum values for $t'\le t$. 
This means that the values of $\omega\left(x,y,t\right)$
at time $t$ are not affected by the evolution of the function
$\omega^*\left(x,y,t\right)$ after time $t$.

\subsection{Outward directed Lorentz force - $\mu$}
\label{muderivation}
Next, we focus on the z-component of the Lorentz force
since a large value indicates which
active regions favour the ejection of magnetic flux ropes. In each of the simulations,
the z-component of the Lorentz force at the lower boundary shows
a complex distribution (Fig.\ref{bzeruptive}).
However, the photospheric Lorentz force represents an incomplete description of the forces
that are acting as the equilibrium of magnetic structures results from an interplay of forces exerted at different heights in the atmosphere.
Therefore, we compute the integral of the vertical component of the Lorentz force along the z-direction ($I_{LFZ}$)
\begin{equation}
I_{LFZ}\left(x,y,t\right)=\int_{z=0}^{z=z_{max}} LF_{z}\left(x,y,z,t\right) dz.
\label{lfzintegral}
\end{equation}
\begin{figure}
\centering
\includegraphics[scale=0.25]{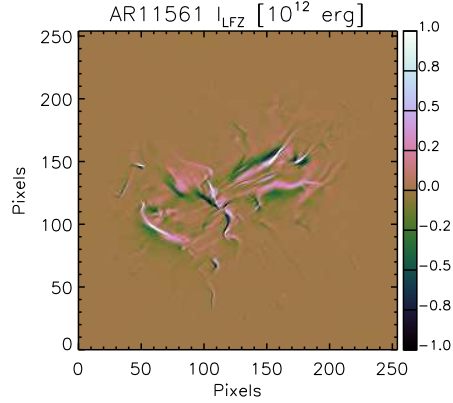}

\caption{Map of the integral of the z-component of Lorentz force along the z direction $I_{LFZ}\left(x,y,t\right)$ near the time of eruption for AR11561.
}
\label{lfzint}
\end{figure}
Fig.\ref{lfzint} shows the distribution of $I_{LFZ}$ at the time of the observed eruption for AR11561.
Typically, we find that the function $I_{LFZ}$ exhibits highly localised positive and negative values.
The lower atmosphere below an arbitrary height of 5 Mm usually contributes more than 75\% to $I_{LFZ}$. 
However, there are many extended spatial locations where this contribution is smaller and more than $25\%$ of the Lorentz force integral originates from heights greater than $5$Mm.


To consider the vertical forces acting on a magnetic flux rope,
we compute the average of the function $I_{LFZ}\left(x,y,t\right)$ 
over a moving circular mask $C^{(x_c,y_c)}_{0.7Mm}$ of radius $0.7$ $Mm$ centred in $(x_c,y_c)$, which we define as
\begin{equation}
\mu^*\left(x,y,t\right)=\frac{\int_{C^{(x_c,y_c)}_{0.7Mm}}^{ } I_{LFZ}\left(x',y',t\right) dx' dy'}{ \pi 0.7^2}.
\label{eqmu}
\end{equation}
The value of $\mu^*\left(x,y,t\right)$ may vary in time at a given location and
can change sign within the same polarities.
We also find that close to or beneath magnetic flux ropes, both positive or negative values of $\mu^*\left(x,y,t\right)$ can occur.
Positive values of $\mu^*\left(x,y,t\right)$ at a specific location
do not necessarily suggest an ejection has occurred, as it is possible that
an outward directed Lorentz force is balanced
by the restoring force of the overlying magnetic field.
However, on average, an outwardly directed Lorentz force is a necessary condition for a flux rope ejection.
A map of $\mu^*\left(x,y,t\right)$ is not shown as it is very similar to Fig.\ref{lfzint}.

For this quantity we also derive $\mu\left(x,y,t\right)$ from $\mu^*\left(x,y,t\right)$
\begin{equation}
\mu\left(x,y,t\right)=\frac{\mu^*\left(x,y,t\right)-min(\mu^*\left(x,y,t'\le t\right))}{max(\mu^*\left(x,y,t'\le t\right))-min(\mu^*\left(x,y,t'\le t\right))}
\label{mrenorm}
\end{equation}
as explained in Sect.\ref{omegaderivation}.

\subsection{Lorentz force heterogeneity - $\sigma$}
\label{sigmaderivation}
For the final quantity in the construction of the metric,
we are interested in identifying locations where the overlying magnetic field
does not balance new positive forces generated at the lower boundary during the evolution.
Therefore, we compute the mean quadratic departure from the average Lorentz force, which is computed using the same circular mask as Eq.\ref{eqmu}.
\begin{equation}
\sigma^*\left(x,y,t\right)=\frac{\int_{C^{(x_c,y_c)}_{0.7Mm}}^{ } \sqrt{\left[I_{LFZ}\left(x',y',t\right)-\mu\left(x,y,t\right)\right]^2} dx' dy'}{ \pi 0.7^2}.
\label{eq3}
\end{equation}
The quantity $\sigma^*\left(x,y,t\right)$ is a measure of how heterogeneous, $I_{LFZ}\left(x,y,t\right)$, 
is in the area under investigation.
We find that there may or may not be a simple correlation between the distributions of $\sigma^*\left(x,y,t\right)$ and $\mu^*\left(x,y,t\right)$.
However, there are spatial locations where both functions have high values.
At these locations, the integral of the Lorentz force is positive and heterogeneous, indicating
that within these locations the Lorentz force is significantly higher and lower than its mean value.
\begin{figure}
\centering
\includegraphics[scale=0.25]{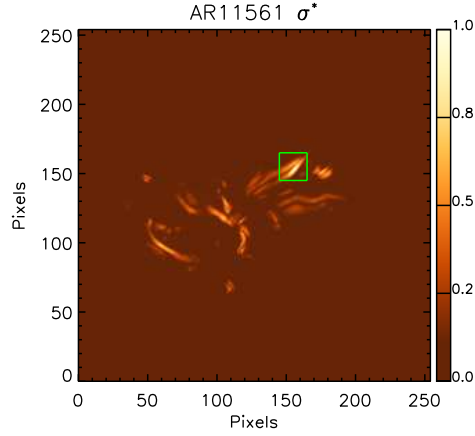}
\caption{Map of the function $\sigma^*$ (Eq.\ref{eq3}) for AR11561 at the time
when an eruption is observed. The green square identifies the location
where the maximum of the distribution is located.
Values are normalised to the maximum of the function at this time.}
\label{sigmaAR11561}
\end{figure}
Fig.\ref{sigmaAR11561} shows a map of the function $\sigma^*\left(x,y,t\right)$ for the eruptive active region AR11561 at the time of the eruption.
We find that only a few elongated structures of high $\sigma^*\left(x,y,t\right)$ are present in the domain,
whereas in most of the active region $\sigma^*\left(x,y,t\right)$ remains rather low compared to its maximum value.
There is one particular structure, which is highlighted by the green square in Fig.\ref{sigmaAR11561},
that shows a large value of $\sigma^*\left(x,y,t\right)$.

Finally, we also apply a normalisation to $\sigma^*\left(x,y,t\right)$ to derive $\sigma\left(x,y,t\right)$, defined as
\begin{equation}
\sigma\left(x,y,t\right)=\frac{\sigma^*\left(x,y,t\right)-min(\sigma^*\left(x,y,t'\le t\right))}{max(\sigma^*\left(x,y,t'\le t\right))-min(\sigma^*\left(x,y,t'\le t\right))}.
\label{srenorm}
\end{equation}

\subsection{Eruption metric - $\zeta$}

The normalisation of $\omega\left(x,y,t\right)$, $\mu\left(x,y,t\right)$, and $\sigma\left(x,y,t\right)$ allows for the comparison of the eruption metric zeta between different active regions.
Moreover, the normalised quantities plateau their values when the non-normalised functions increase in time.

By comparing Figures \ref{omegaAR11561}, \ref{lfzint}, and \ref{sigmaAR11561}, it is apparent that for eruptive active regions
the spatial locations over which $\omega$ shows higher values
includes the corresponding locations where either $\mu$ or $\sigma$ show high values.
We also find that each individual function can have a value close to 1,
however this rarely happens simultaneously for all three functions.
The same conclusions on the spatial distribution of $\omega\left(x,y,t\right)$, $\mu\left(x,y,t\right)$, and $\sigma\left(x,y,t\right)$ can be drawn for the active regions where no eruptions are found.
As anticipated, the newly introduced eruption metric $\zeta$ (Eq.\ref{zeta})
combines the information from $\omega$, $\mu$, and $\sigma$,
is bounded between 0 and 1, and
is the product of three normalised quantities that are functions of space and time.
For consistence in notation, we define $\zeta\left(x,y,t\right)$ as
\begin{equation}
\zeta\left(x,y,t\right)=\omega\left(x,y,t\right)\mu\left(x,y,t\right)\sigma\left(x,y,t\right).
\label{zeta2dt}
\end{equation}

\begin{figure}
\centering
\includegraphics[scale=0.25]{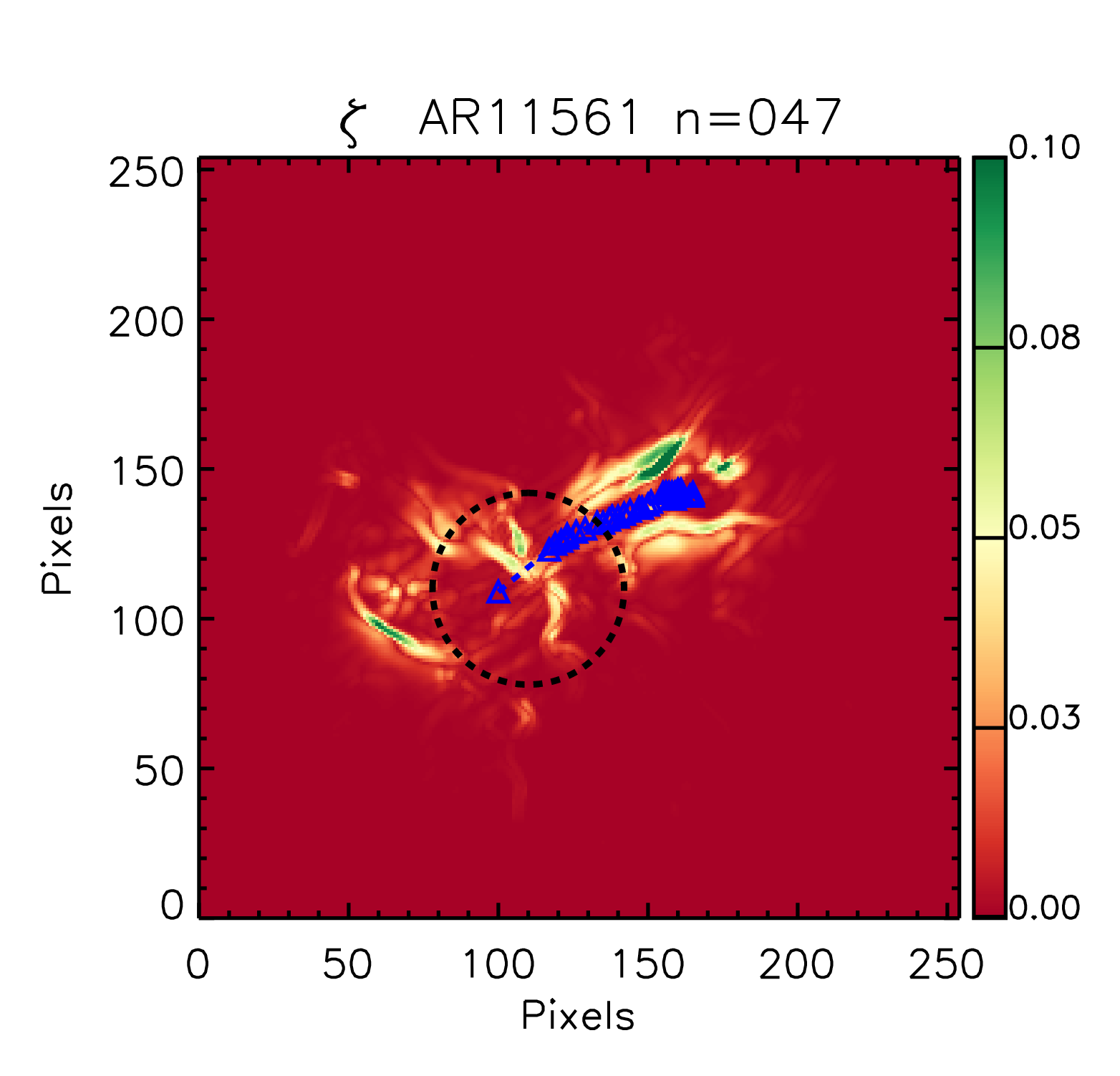}
\includegraphics[scale=0.25]{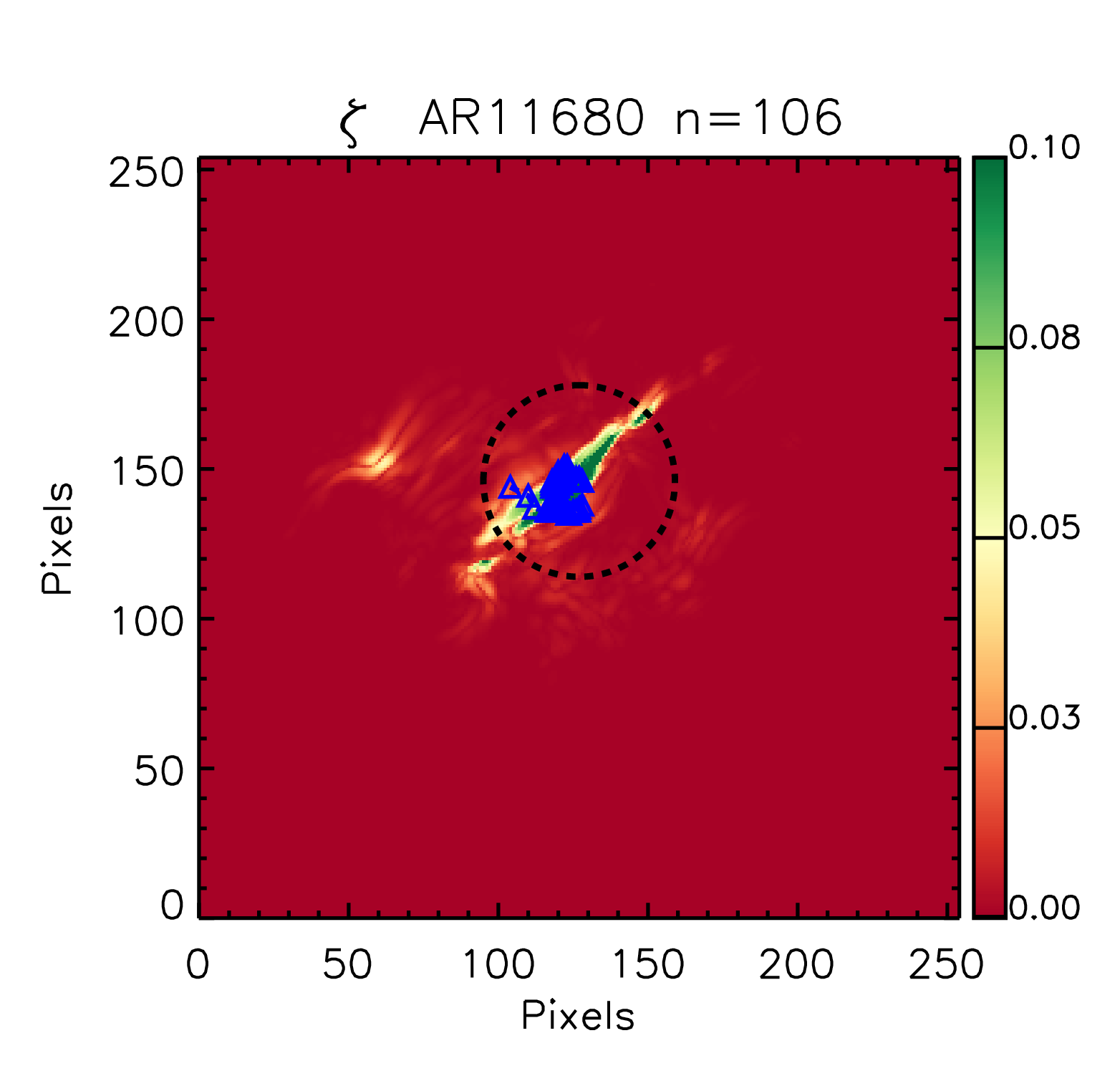}
\includegraphics[scale=0.25]{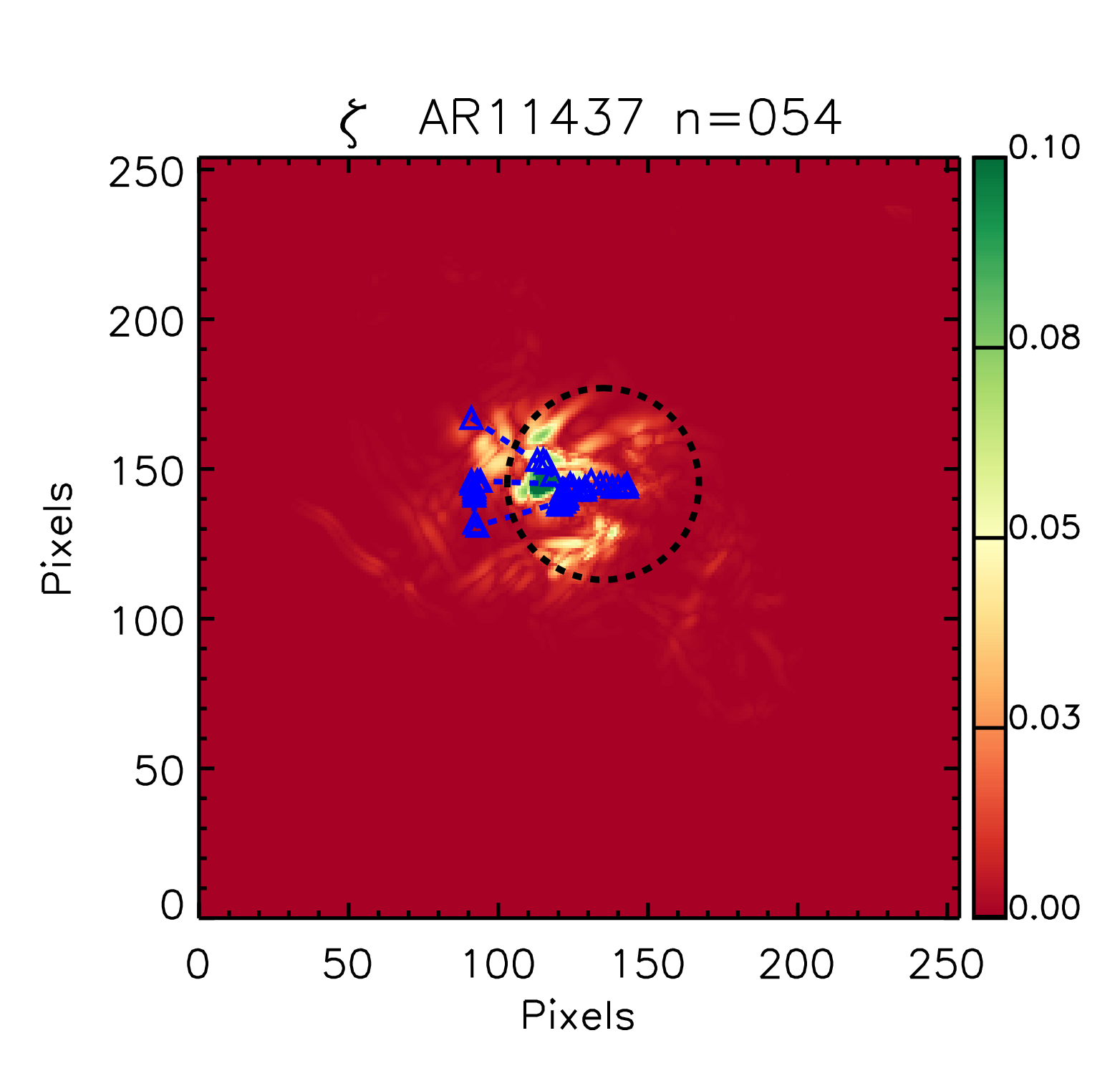}
\includegraphics[scale=0.25]{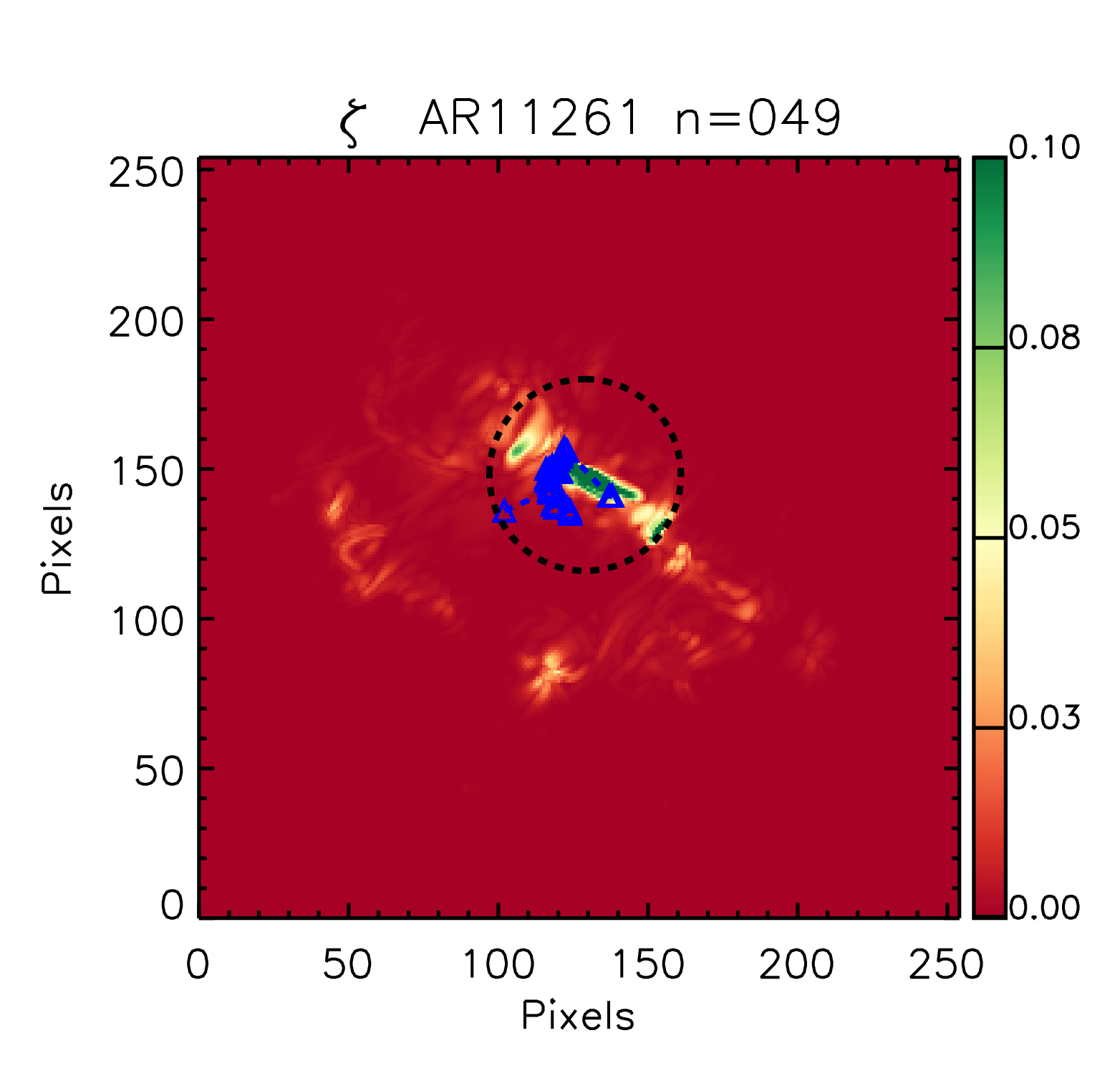}
\includegraphics[scale=0.25]{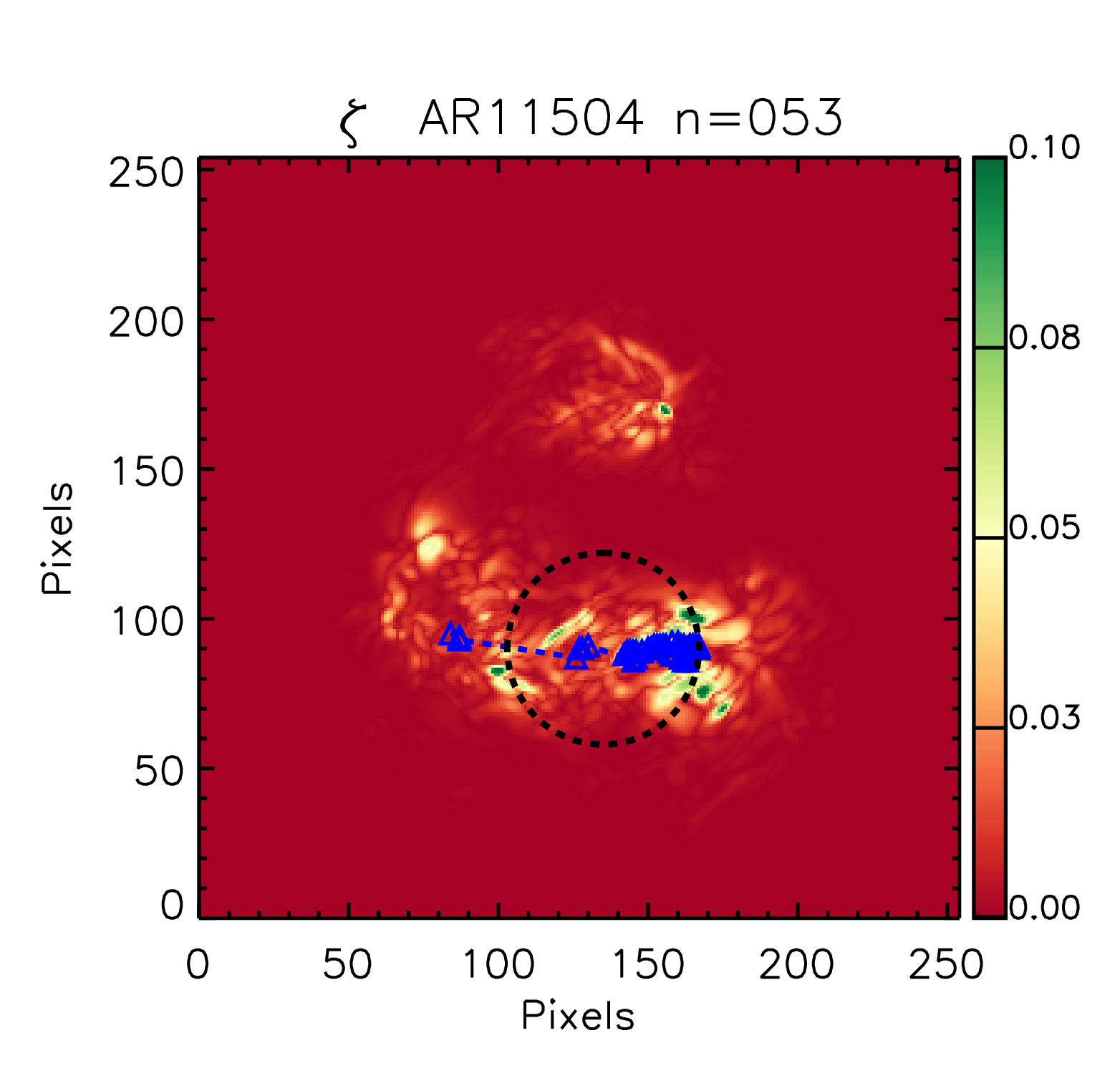}
\caption{Maps of $\zeta\left(x,y,t\right)$ near the time of to the n-th magnetogram when the eruption is osberved for the eruptive active regions of our set.
The black dashed circles identify where the eruption has been located in 
observational studies and the blue triangles connected by the blue dashed line shows
the location of the maximum value of $\zeta$ at different times.}
\label{armapsejectioneruptive}
\end{figure}

\begin{figure}
\centering
\includegraphics[scale=0.25]{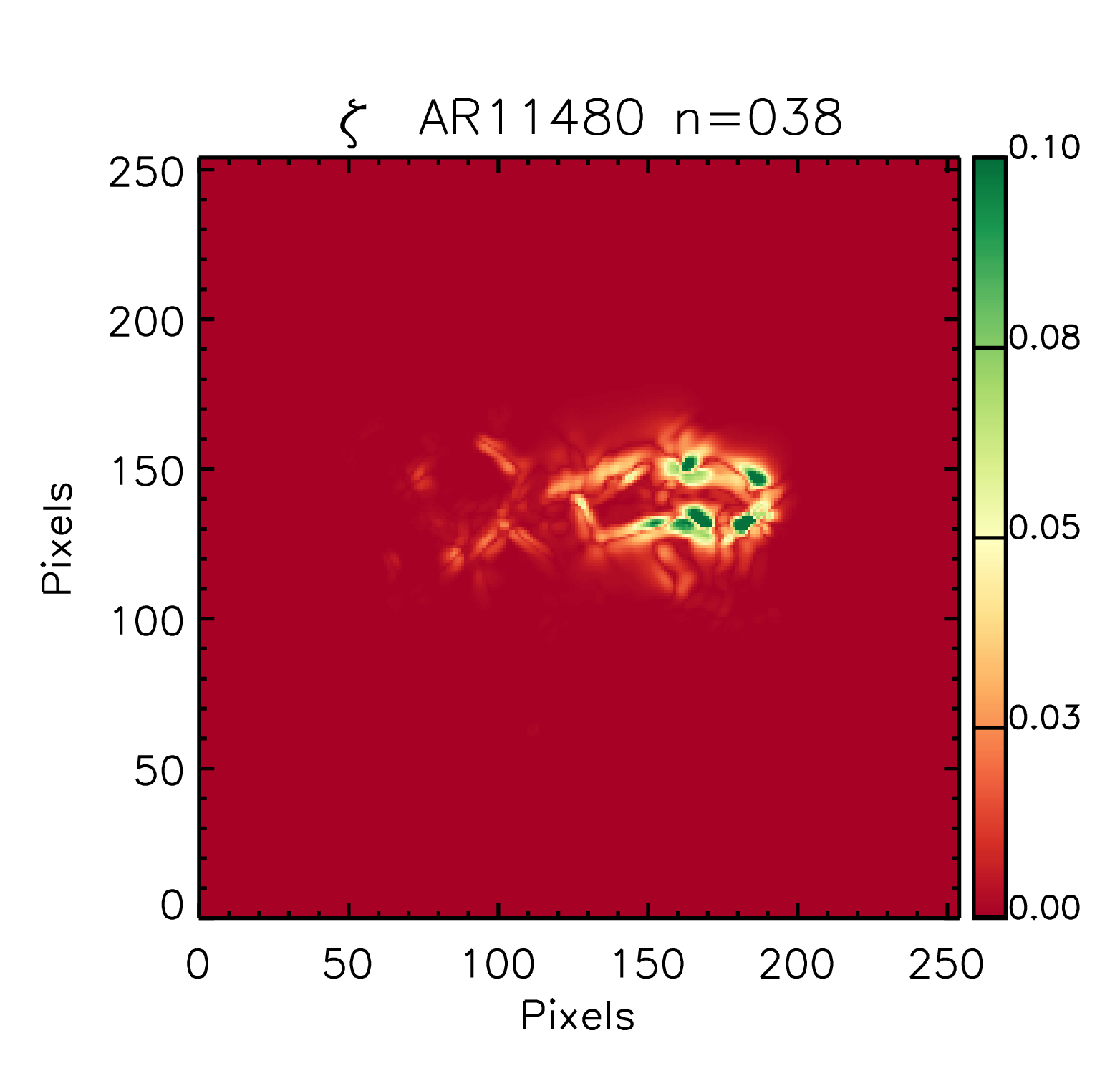}
\includegraphics[scale=0.25]{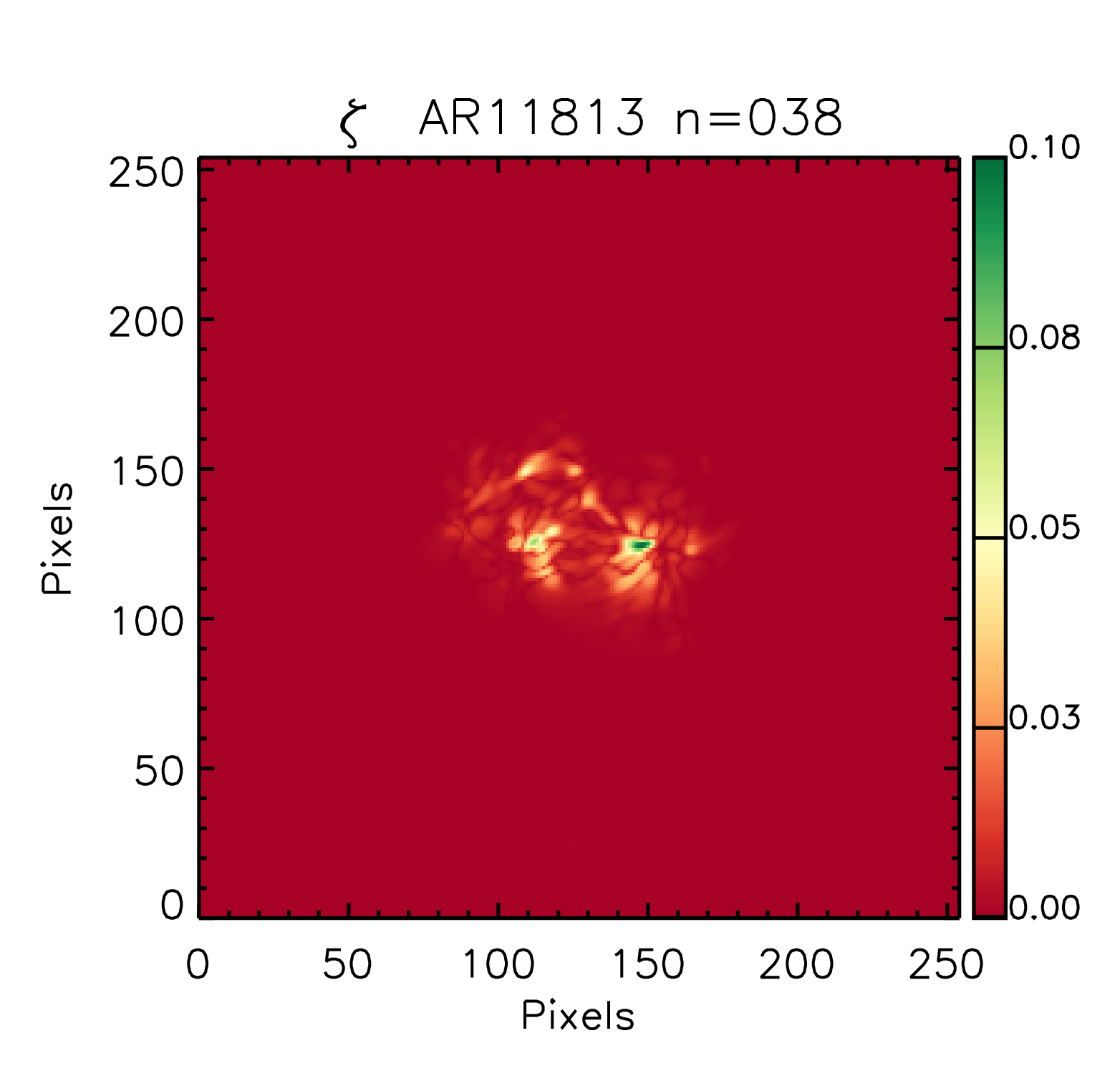}
\includegraphics[scale=0.25]{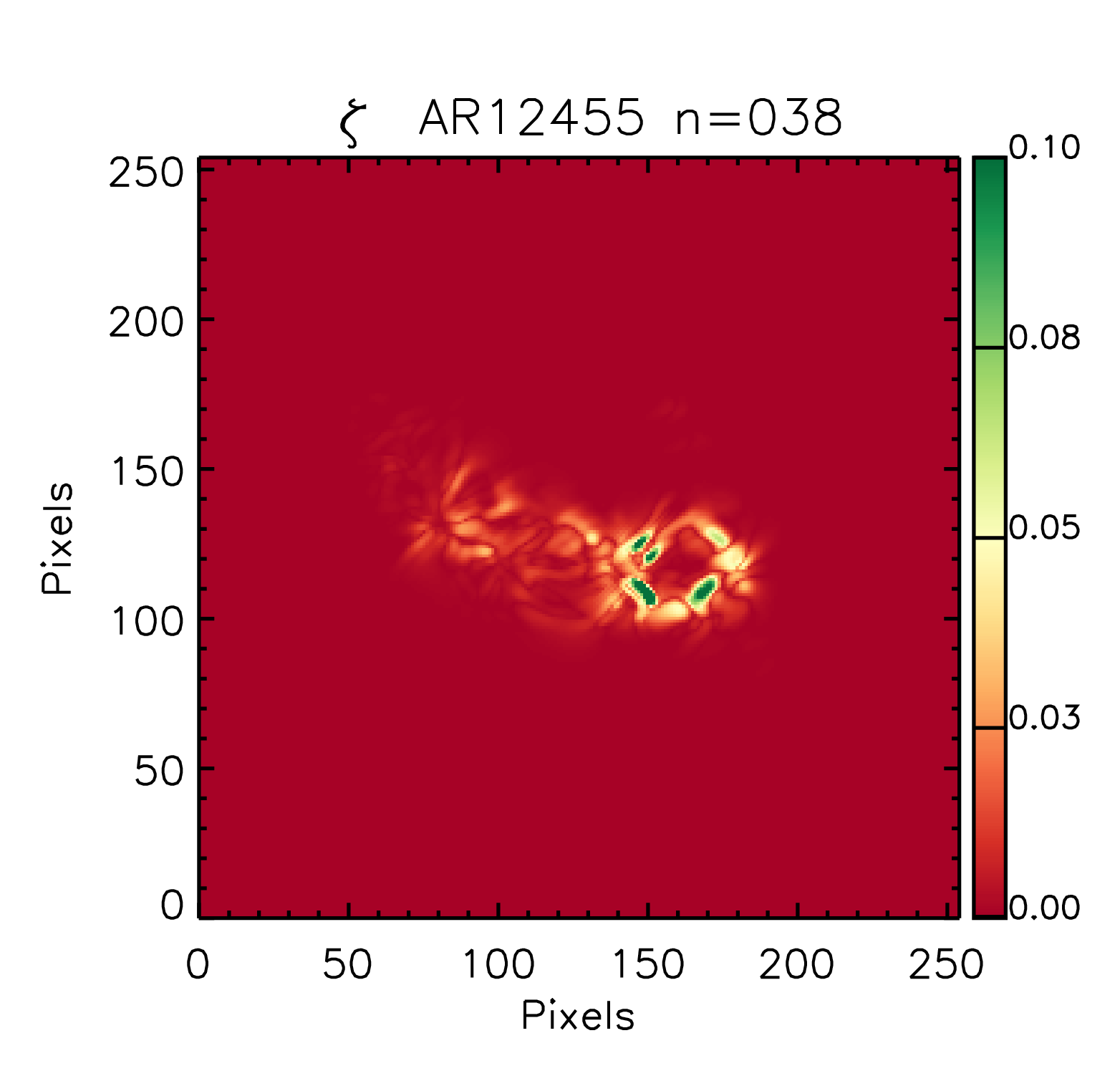}
\caption{Maps of $\zeta\left(x,y,t\right)$
for the non-eruptive active regions taken at the time of to the n-th magnetogram of maximum $\zeta\left(x,y,t\right)$.}
\label{armapsejectionnoneruptive}
\end{figure}

Fig.\ref{armapsejectioneruptive} shows the maps of $\zeta\left(x,y,t\right)$
for the five eruptive active regions in this study, at the time of 
the observed eruption.
Over the majority of the domain we find that the value of $\zeta\left(x,y,t\right)$ is generally close to 0, except at a few locations where it takes a value close to $0.1$.
The black dashed circles identify the origin of the eruptions as observed by previous studies
\citep{Rodkin2017,Yardley2018b,James2018},
while the blue triangles identify the maximum value of $\zeta\left(x,y,t\right)$ at different times in the active region evolution.
We find that the location of the maximum value of $\zeta\left(x,y,t\right)$ usually matches the location of the eruption in the observations.
The match is particularly good for active regions where
the eruption was due to a filament eruption
(AR11680, AR11261, AR11504),
for example the eruption of a large filament that
was associated with the internal polarity inversion line of AR11680.
The path of the filament matches the location of high $\zeta\left(x,y,t\right)$
values in the magnetofrictional simulation of the active region.
In four out of five cases, the eruption occurs at the location of maximum $\zeta\left(x,y,t\right)$,
however, this is not the case for AR11561. For this active region there is strong flux emergence during which the two magnetic polarities diverge. During this divergence the location of maximum $\zeta\left(x,y,t\right)$ moves with one polarity.
The eruption does not occur at the location of maximum $\zeta\left(x,y,t\right)$, however
the value of $\zeta\left(x,y,t\right)$ is still high at the location of the observed eruption.
Fig.\ref{armapsejectionnoneruptive} shows maps of $\zeta\left(x,y,t\right)$
for the three active regions where no eruption was reported. For each case $\zeta\left(x,y,t\right)$ is shown at the time when it reaches its maximum value.
The values of $\zeta\left(x,y,t\right)$ are in general lower and more localised for the non-eruptive active regions compared to that found for the eruptive active regions.

\subsection{Time evolution of the eruption metric $\zeta$}
\begin{figure}
\centering
\includegraphics[scale=0.40]{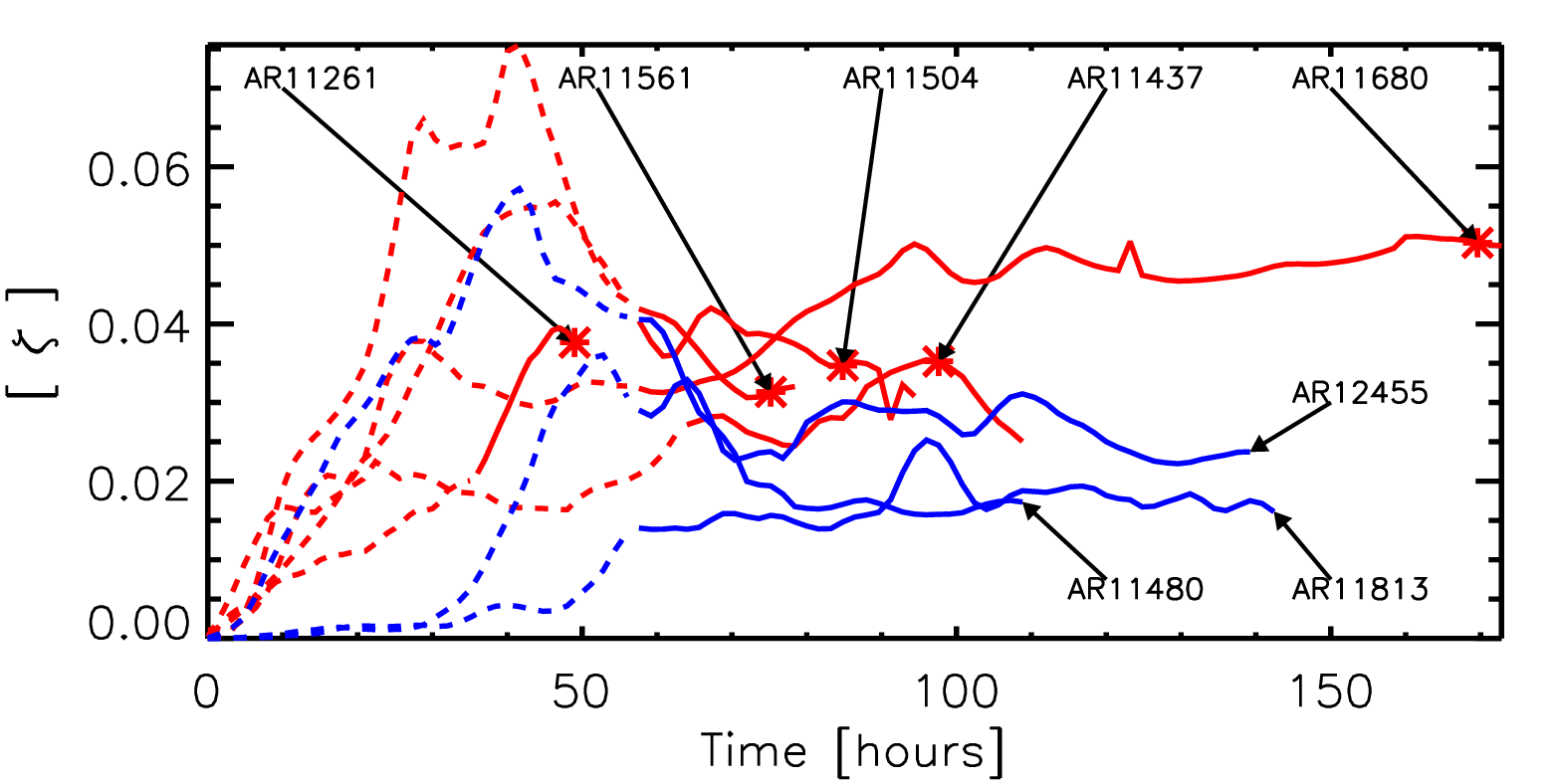}
\caption{Evolution of the maximum value of $\zeta\left(x,y,t\right)$ 
for all active regions in our dataset.
The red curves represent eruptive active regions and the blue curves non-eruptive ones.
Dashed curves represent the first 35 magnetograms in the series.
The asterisks indicate the timings of the eruptions originating from the active regions
as given by the corresponding literature.
Times are reported from the first magnetogram observation for each active region.
}
\label{critevol}
\end{figure}

To distinguish eruptive from non-eruptive active regions,
we carry out a twofold process.
First of all, we compute a spatial average of the eruption metric $\zeta\left(x,y,t\right)$
over a square of size 5.8 Mm to remove local effects.
Next, we consider the maximum of the spatial average and we consider
the evolution of this maximum obtined as a function of time, $\zeta_{max}\left(t\right)$.
Fig.\ref{critevol} shows the resulting evolution of
$\zeta_{max}\left(t\right)$ for each of
the active regions considered here.
The red curves represent eruptive active regions, with the 
red asterisks indicating the time of eruption as seen in the observations, 
and the blue curves represent non-eruptive active regions.
We find that, for all simulations, there is an increase in $\zeta_{max}\left(t\right)$ at the start of the evolution,
which is due to the injection of electric currents from boundary motions.
This leads to the magnetic field configuration departing from its initial potential state.
We estimate that it takes around 35 magnetograms
(between 40 and 55 hours depending on the magnetograms cadence)
for the magnetofrictional simulation to lose memory of the initial potential configuration
and we consider this the ramp-up phase of the magnetofrictional simulations.
After this ramp-up phase,
the system tends to converge to a value of $\zeta_{max}\left(t\right)$ that is significantly different
for eruptive active regions compared to the non-eruptive ones.
Eruptive active regions tend to converge to values between 0.03 and 0.05,
while non-eruptive active regions usually converge to values around 0.02.
However, the evolution of $\zeta_{max}\left(t\right)$
still fluctuates after the initial ramp-up phase.
Some active regions (AR11261, AR11504, AR11437) show an instantaneous decline of $\zeta_{max}\left(t\right)$
post-eruption.
This does not occur for
AR11561 as the magnetogram series ceases post-eruption.
AR11680 shows the highest value of $\zeta_{max}\left(t\right)$ 
for all of the simulations and for an extended period of time.

The present analysis does not provide a unique way 
to link high values of $\zeta_{max}\left(t\right)$ to the likelihood of an eruption
due to two main features.
The first one is that non-eruptive active regions show values of $\zeta_{max}\left(t\right)$ instantaneously higher than eruptive active regions and vice versa, close to the ramp up phase.
The second one is that for the eruptive active regions, the time of the eruption
does not always coincide with the time when $\zeta_{max}\left(t\right)$ is at its maximum.

\begin{figure}
\centering
\includegraphics[scale=0.40]{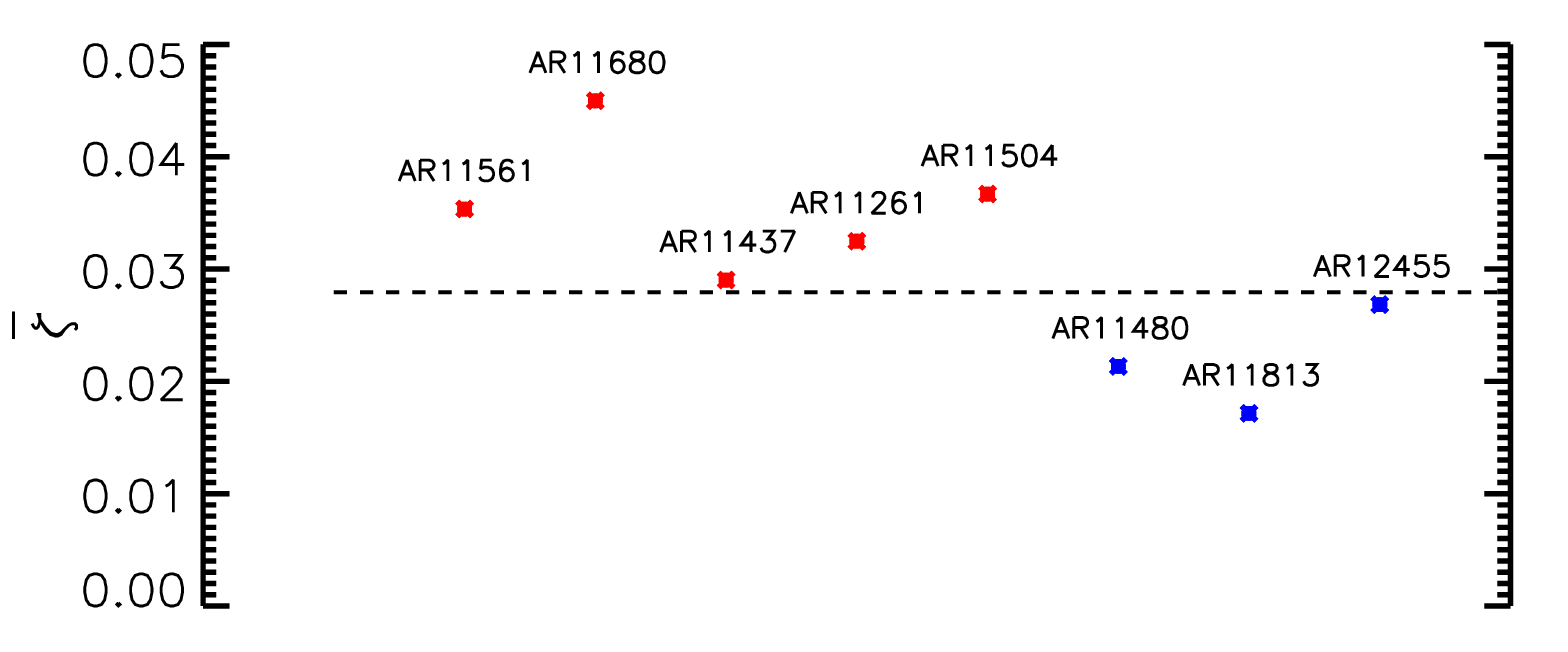}
\caption{The values of $\bar{\zeta}$ for each active region. Red asterisks are the eruptive active regions and blue asterisks represent the non-eruptive ones.
The dashed curve represents the value of $\bar{\zeta_{th}}=0.028$.}
\label{critars}
\end{figure}

However, the values of $\zeta_{max}\left(t\right)$ in eruptive active regions
are generally higher than in non-eruptive active regions.
Therefore, this property can be used to
define a new metric that distinguishes
between eruptive and non-eruptive active regions.
For that reason, we consider $\bar{\zeta}$, the time average of $\zeta_{max}\left(t\right)$ between the end of the ramp-up phase and the final magnetogram for each active region.
For some active regions, the final magnetogram is after the observed eruption time, but this has a marginal effect on the time average.
The time average of $\zeta_{max}\left(t\right)$ is a single number and can therefore be directly associated to the likelyhood that an active region will produce an eruption.
Fig.\ref{critars} shows the value of $\bar{\zeta}$ for each active region.
We find that eruptive active regions in our sample show significantly higher values of $\bar{\zeta}$ and are separated from the non-eruptive active regions.
For purely operational purposes, we compute a threshold of $\bar{\zeta_{th}}=0.028$,
as the average between the maximum value of $\bar{\zeta}$ among the non-eruptive active regions 
and the minimum value of $\bar{\zeta}$ among the eruptive active regions
(dashed horizontal line in Fig.\ref{critars}).
It should be noted, that the scattering in $\bar{\zeta}$ within the populations of eruptive
and non-eruptive active regions is larger
than the minimum difference in $\bar{\zeta}$ between the two populations.
As a consequence, it remains a possibility that if a larger sample of active regions
is considered then the two populations would partially overlap. This will be considered in future
studies along with improvements in the metric.

The present study shows that it is possible to identify a metric and a threshold value based only on the magnetic field configuration that discerns between eruptive and non-eruptive active regions.
Of course, this claim is based on a limited sample of eight active regions that we have analysed, but
an important aspect is that the present analysis makes no specific assumption of the
triggering mechanism of the eruption or the processes that lead to the accumulation and release of free magnetic energy. If future larger sample studies show that these results are robust
the present modelling and analysis technique has allowed us 
to derive a threshold $\bar{\zeta}$ that represents the likelyhood of an eruption occurring in an active region
from a series of magnetogram measurements coupled with a NLFFF evolutionary model. If the robustness
is shown then the technique will have significant operational capacity.

\section{Projection of magnetograms}
\label{prediction}
We now investigate whether the eruption metric can distinguish eruptive from
non-eruptive active regions when we project the magnetic field evolution forward in time, i.e. after a certain time $t_0$ we no longer use the observed magnetograms as the lower boundary condition in the magnetofrictional simulations,
but continue to evolve the lower boundary as discussed below.
In particular, we project forward in time the magnetic field evolution of each active region to quantify the corresponding projected value of $\bar{\zeta}$.


\subsection{Projected Active Region Evolution}
\label{evolutionextrapolation}
To evolve the surface and coronal fields beyond $t_0$, we project the evolution of the magnetograms
from $t=t_0$ to the time of the final magnetogram in the observed time sequence $t=t_f$. The method of projection is now described. 
Let $\vec{A}_{pt}= (A_x, A_y)$ be the vector potential at the photospheric boundary that
can be integrated from $B_z$.
The observed magnetograms ($B_{z}$) and derived vector potential, $\vec{A}_{pt}$, are used to
drive the evolution of the coronal field until time $t_0$.
We assume that the electric field at the lower boundary remains constant from $t_0$ to $t_f$:
\begin{equation}
\frac{\partial \vec{A}_{pt}\left(t > t_0 \right)}{\partial t}=\frac{\vec{A}_{pt}\left(t_0 \right)-\vec{A}_{pt}\left(t_0-\Delta t \right)}{\Delta t} = \vec{E_{pt}},
\label{extrapolateA}
\end{equation}
where $\vec{E_{pt}}$ is the electric field deduced from the last two observed magnetograms used in the simulation at $t=t_0$ and $t=t_0-\Delta t$ respectively.
The time cadence of the magnetograms is either $\Delta t=96$ $mins$ or $\Delta t=60$ $mins$.
This means that the vector potential at the lower buondary after $t_0$ is constructed by using
\begin{equation}
\vec{A}_{pt}\left(t > t_0 \right)=\vec{A}_{pt}\left(t_0 \right)+\left(t-t_0\right)\vec{E_{pt}}
\label{extrapolatedA}
\end{equation}

While $\vec{E_{pt}}$ remains constant in time, it varies spatially from one pixel to the next.
This is equivalent to observing an active region until a given time, $t=t_0$, after which we project its future evolution by using the most recent information available from the magnetogram time series.
In this way, we have a hybrid simulation based on both observed and projected magnetograms.
We can perform the same analysis as in Sect.\ref{criteria} to find the evolution of $\zeta_{max}\left(t\right)$ and
the value of $\bar{\zeta}$ associated with the hybrid simulation.

\subsection{Identifying eruptive active regions}
\label{predictionresults}

To understand how the evolution of $\zeta_{max}\left(t\right)$ is affected by the projection of the magnetograms we vary the time which we stop using the observed magnetograms as the lower boundary conditions of the simulation.
To simplify the comparison we choose to terminate all simulations at $t=t_f$.
The simulations with $t_0=t_f$ are described in Sec.\ref{model} where all available observational
data are used to reproduce the coronal evolution of the active region.
For each active region, we run 19 simulations with projected magnetograms,
i.e. varying from $t_0=t_f-19$ $\Delta t$ to $t_0=t_f-$ $\Delta t$.

\begin{figure}
\centering

\includegraphics[scale=0.18]{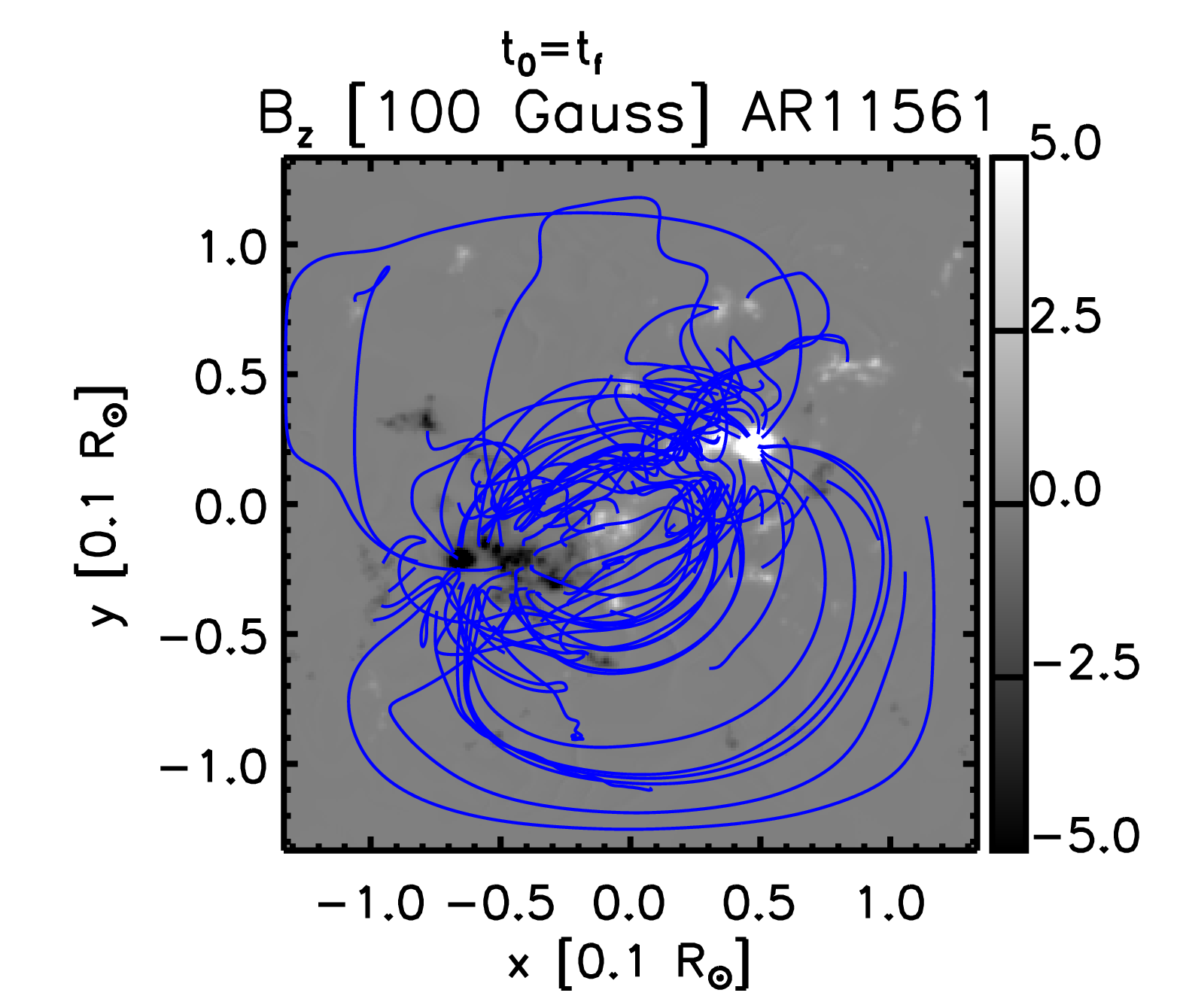}
\includegraphics[scale=0.18]{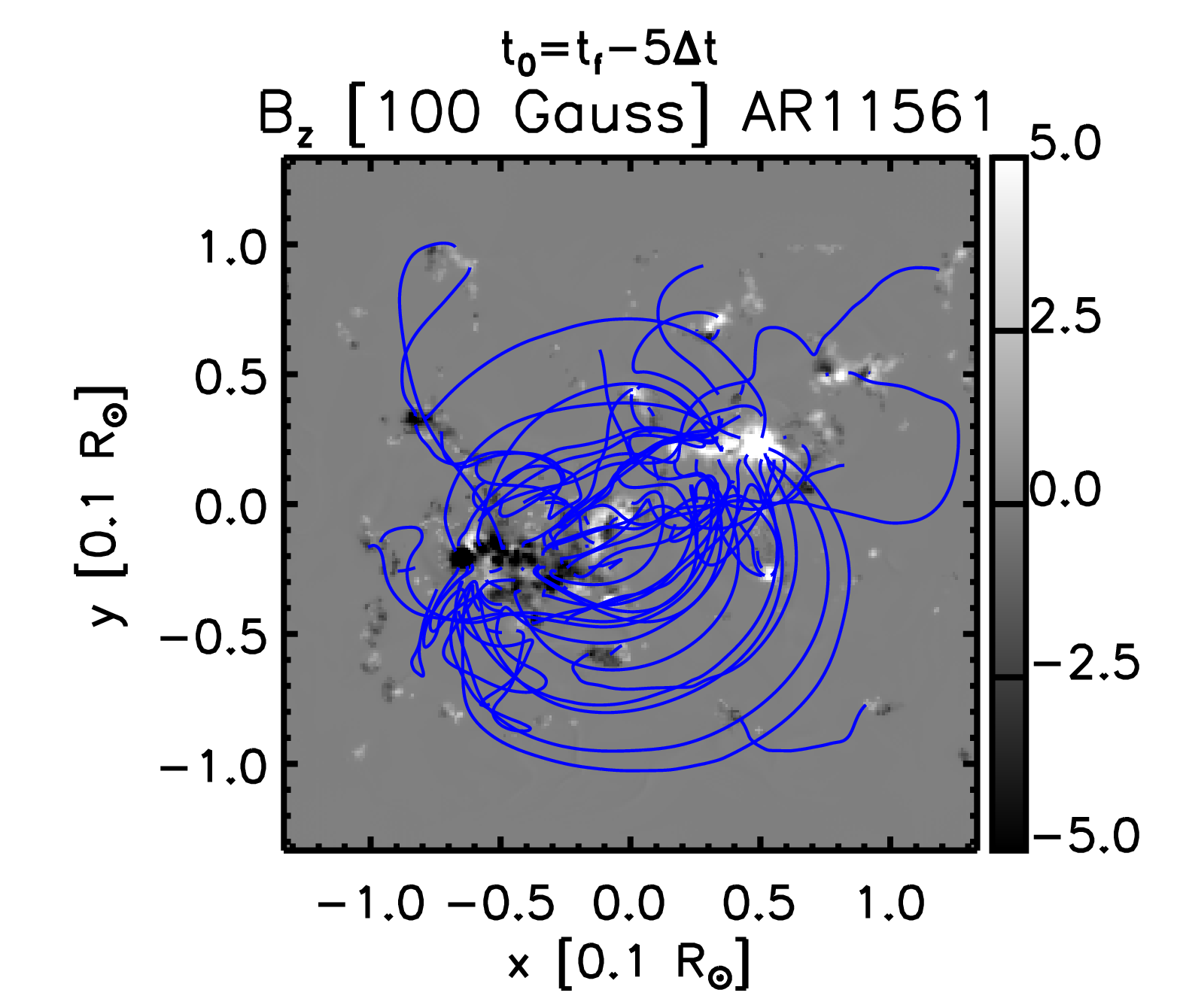}
\includegraphics[scale=0.18]{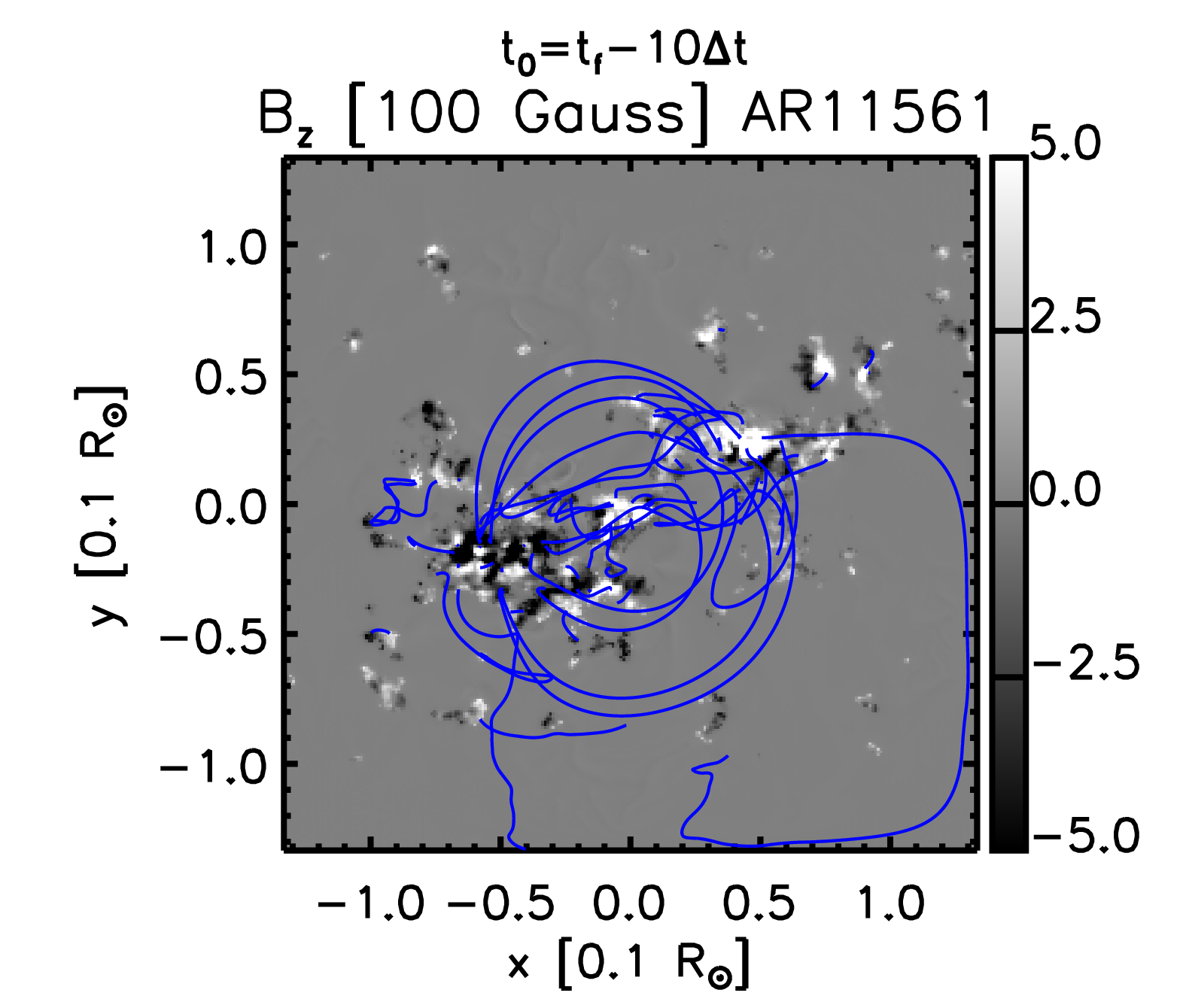}

\includegraphics[scale=0.17]{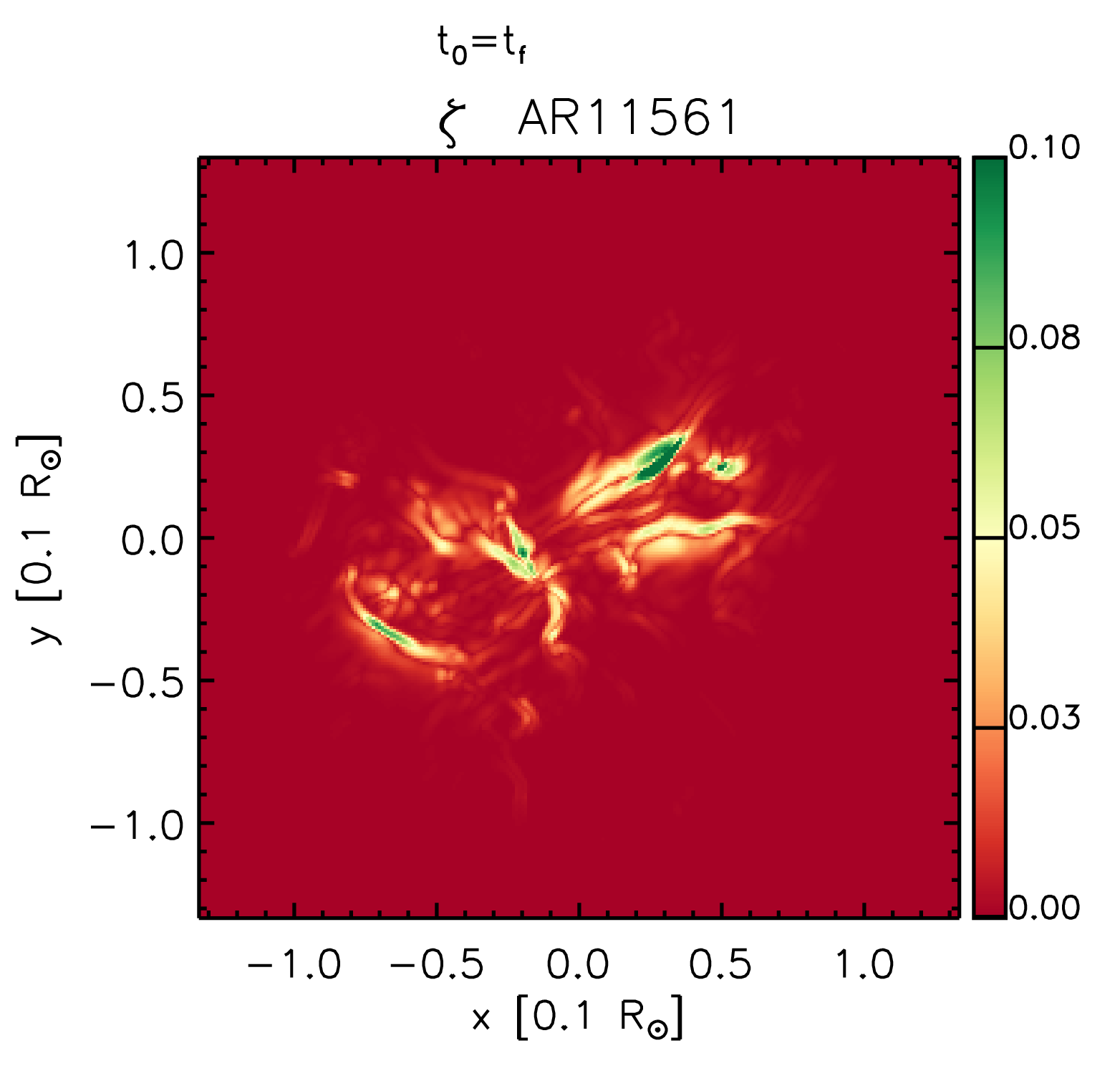}
\includegraphics[scale=0.17]{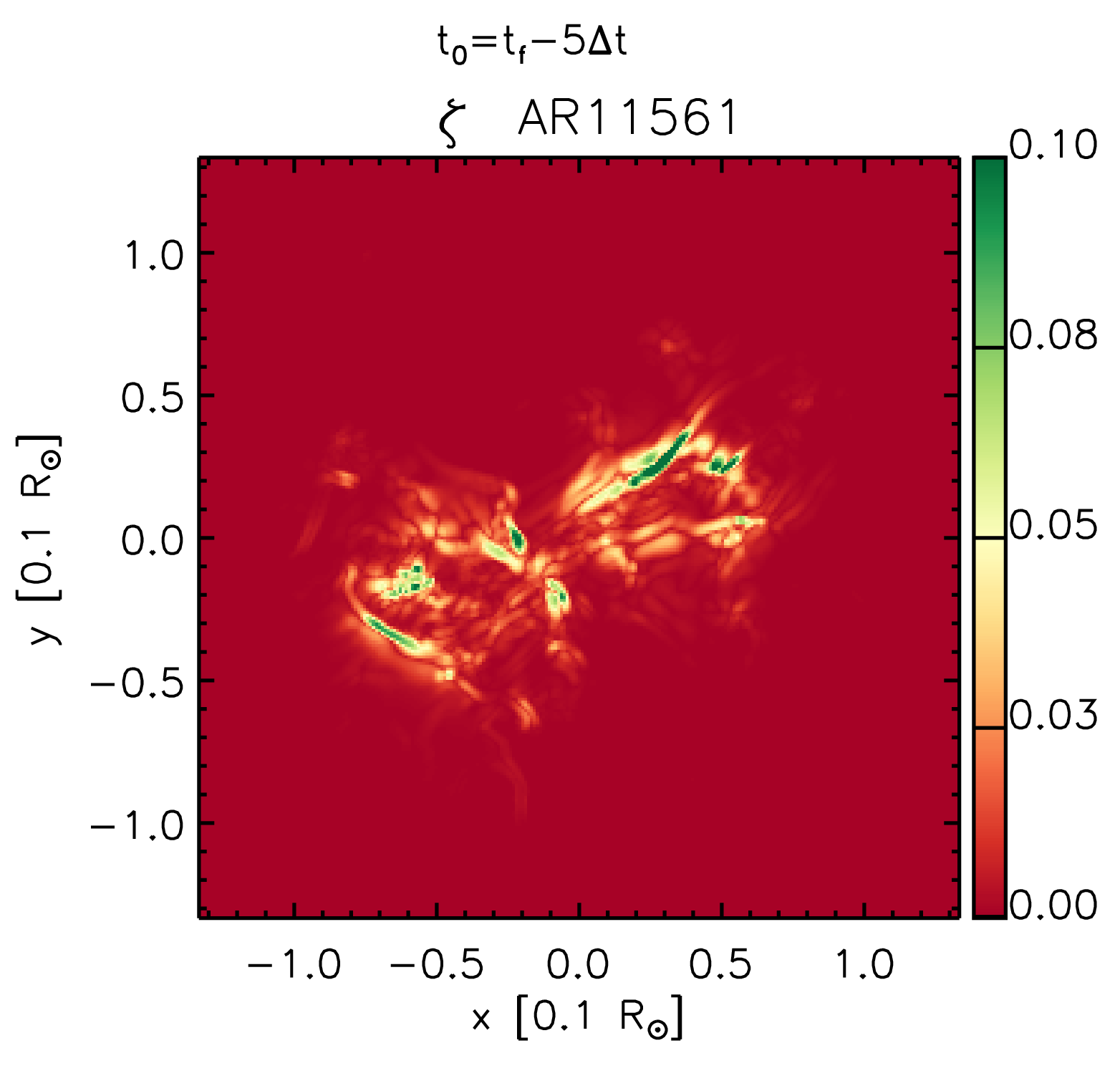}
\includegraphics[scale=0.17]{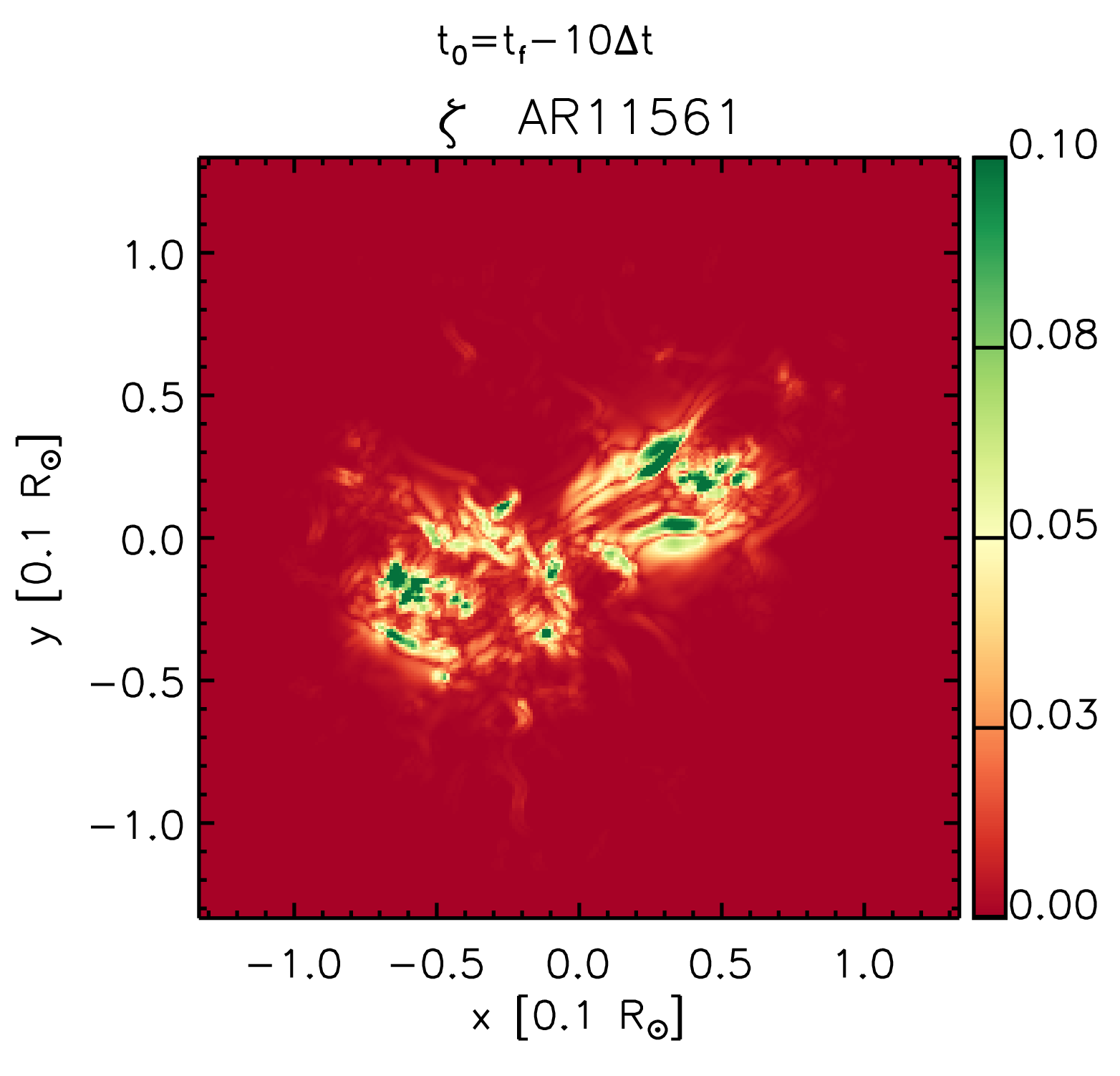}

\caption{Maps of the photospheric $B_z$ with representative magnetic field curves overplotted
at $t=t_f$ for three simulations of AR11561
(left) the reference simulation, (centre) the simulation with $t_0=t_f-5\Delta t$  and (right) $t_0=t_f-10\Delta t$.}
\label{comparisonB_AR11561}
\end{figure}

Fig.\ref{comparisonB_AR11561} shows
maps of the surface magnetic field $B_z$ for the final state ($t=t_f$) of the simulations
where $t_0=t_f$ (all observations are used), $t_0=t_f-5\Delta t$, and $t_0=t_f-10\Delta t$ for AR11561.
We find that the simulation with the projected magnetograms from $t_0=t_f-5\Delta t$
reproduces the majority of the magnetic field features (loops and surface magnetic field distribution)
found in the final magnetic configuration of the full data simulation ($t_0=t_f$).
In contrast, the projected simulations from $t_0=t_f-10\Delta t$ are visibly different from the full observed data simulation.
However, these differences are mostly found along the boundaries
of the flux concentrations and within weaker magnetic field locations.
Due to this, they do not significantly affect the overall connectivity of the coronal field of the active region.

\begin{figure}
\centering
\includegraphics[scale=0.28]{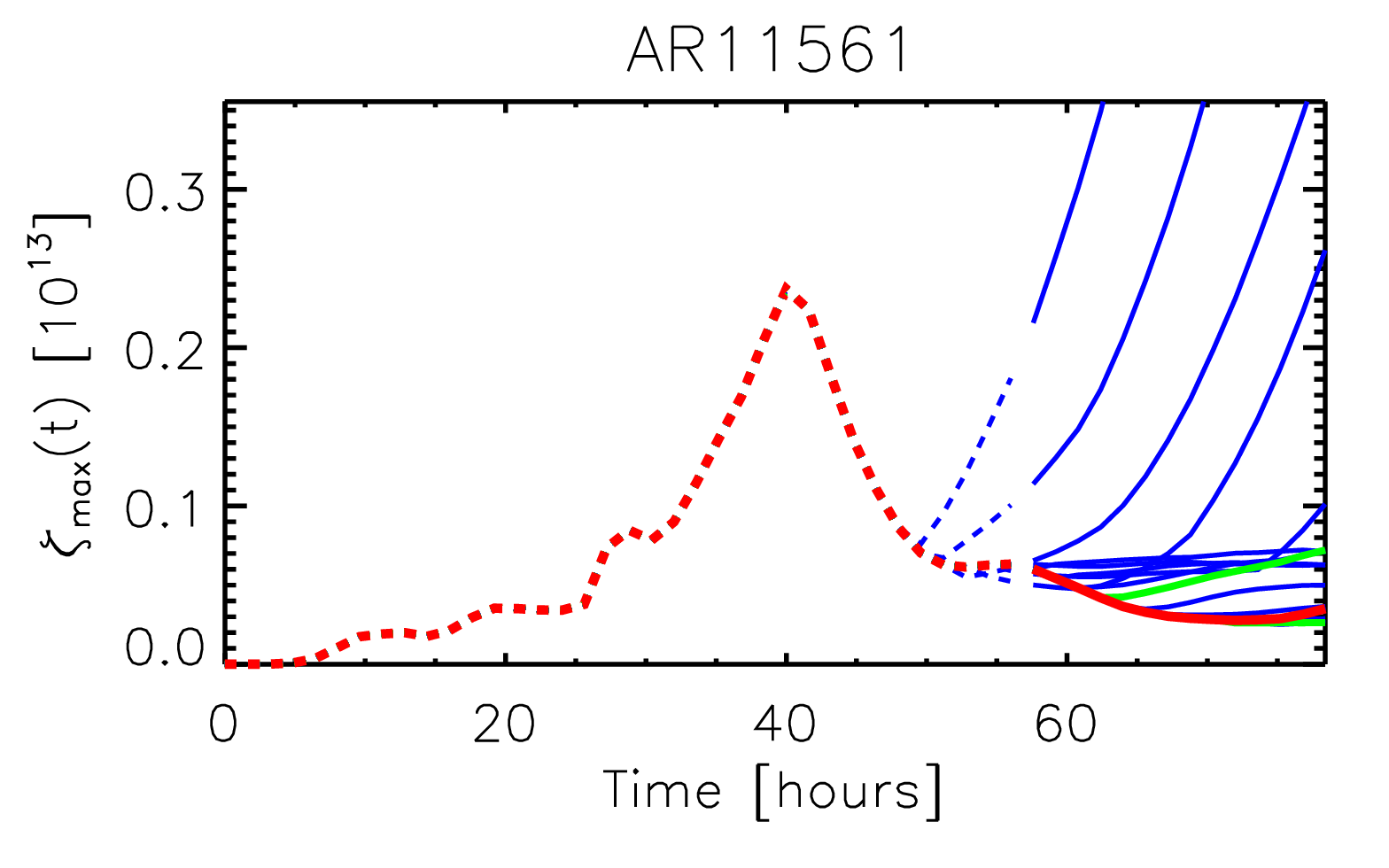}
\includegraphics[scale=0.28]{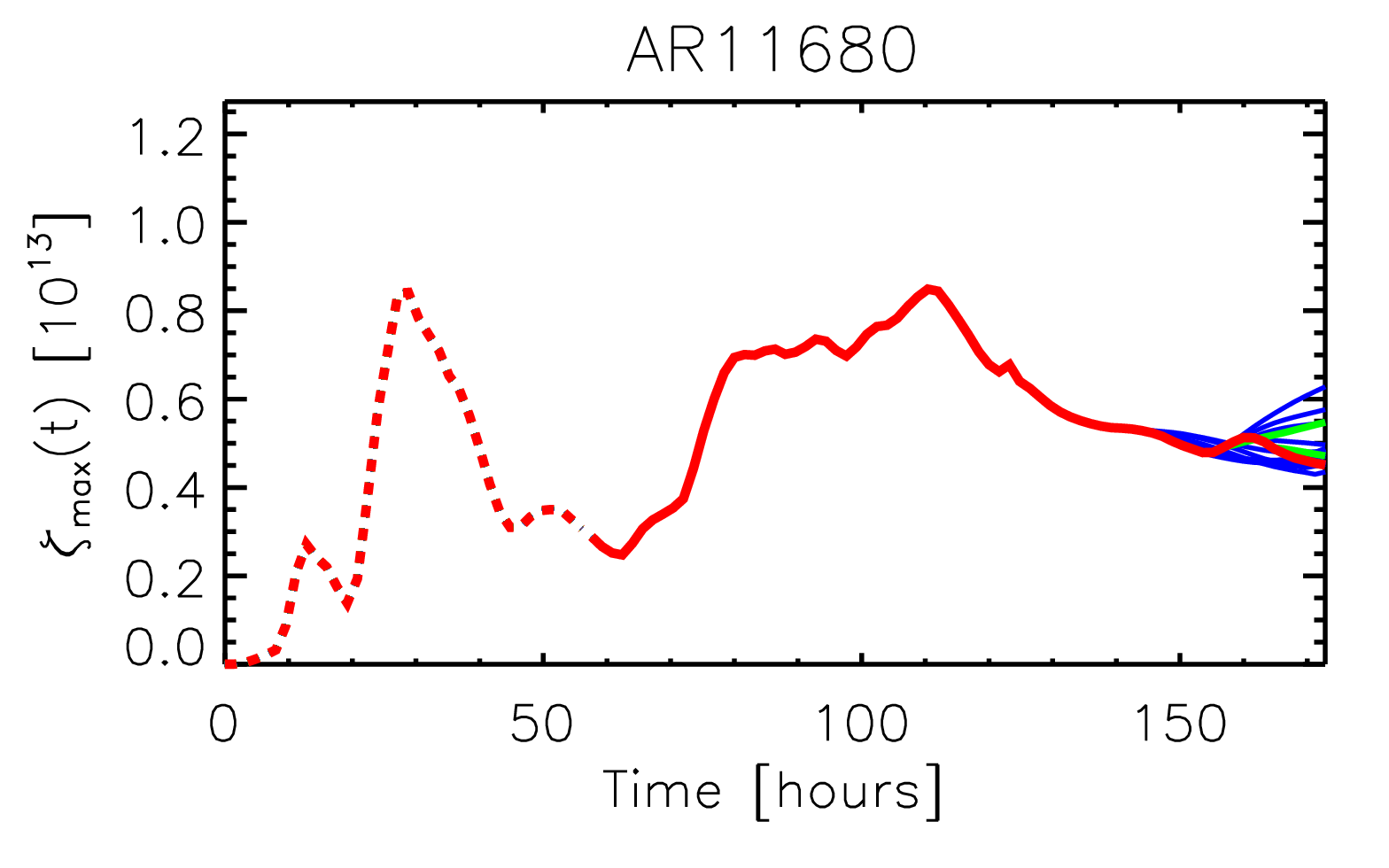}
\includegraphics[scale=0.28]{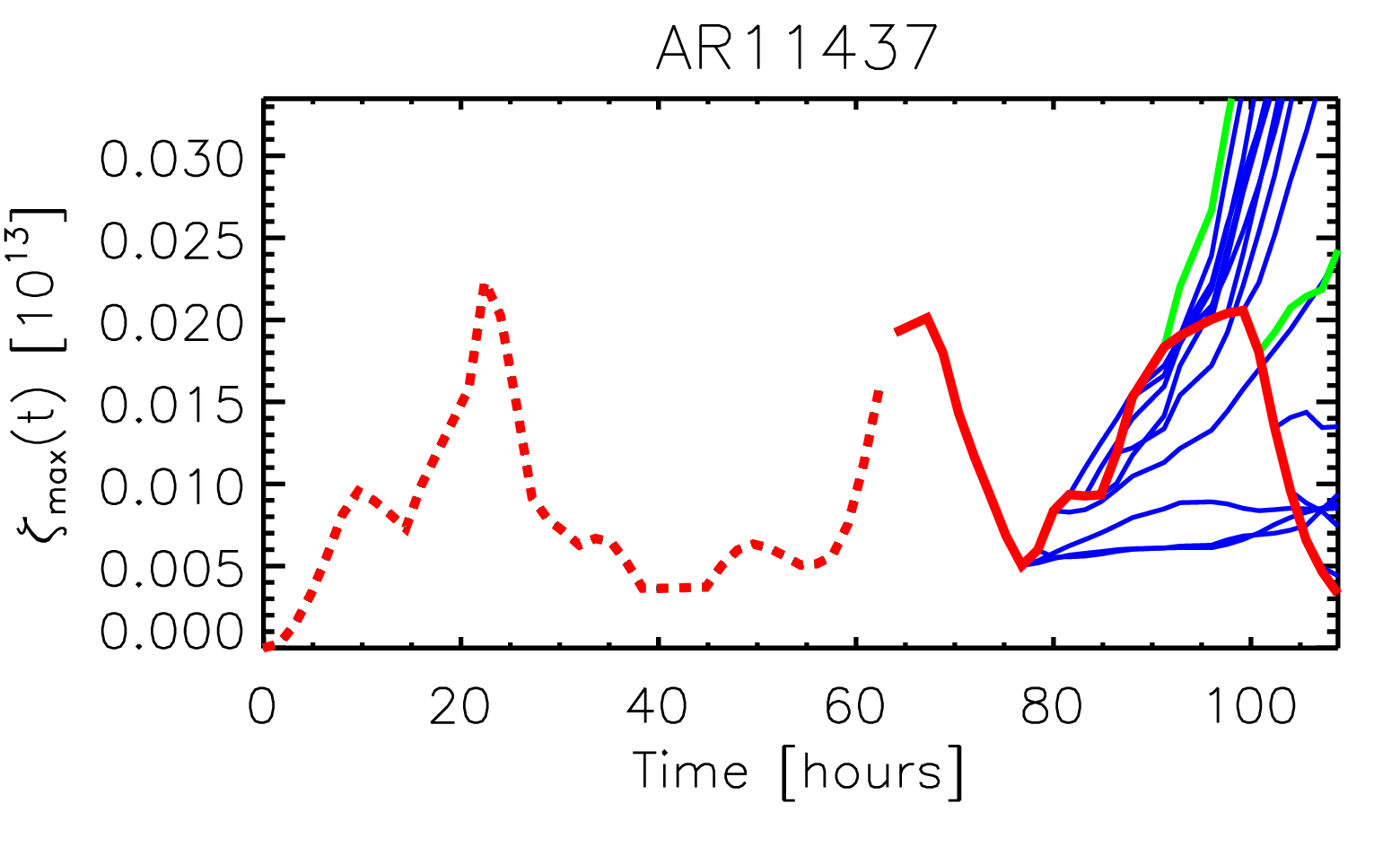}
\includegraphics[scale=0.28]{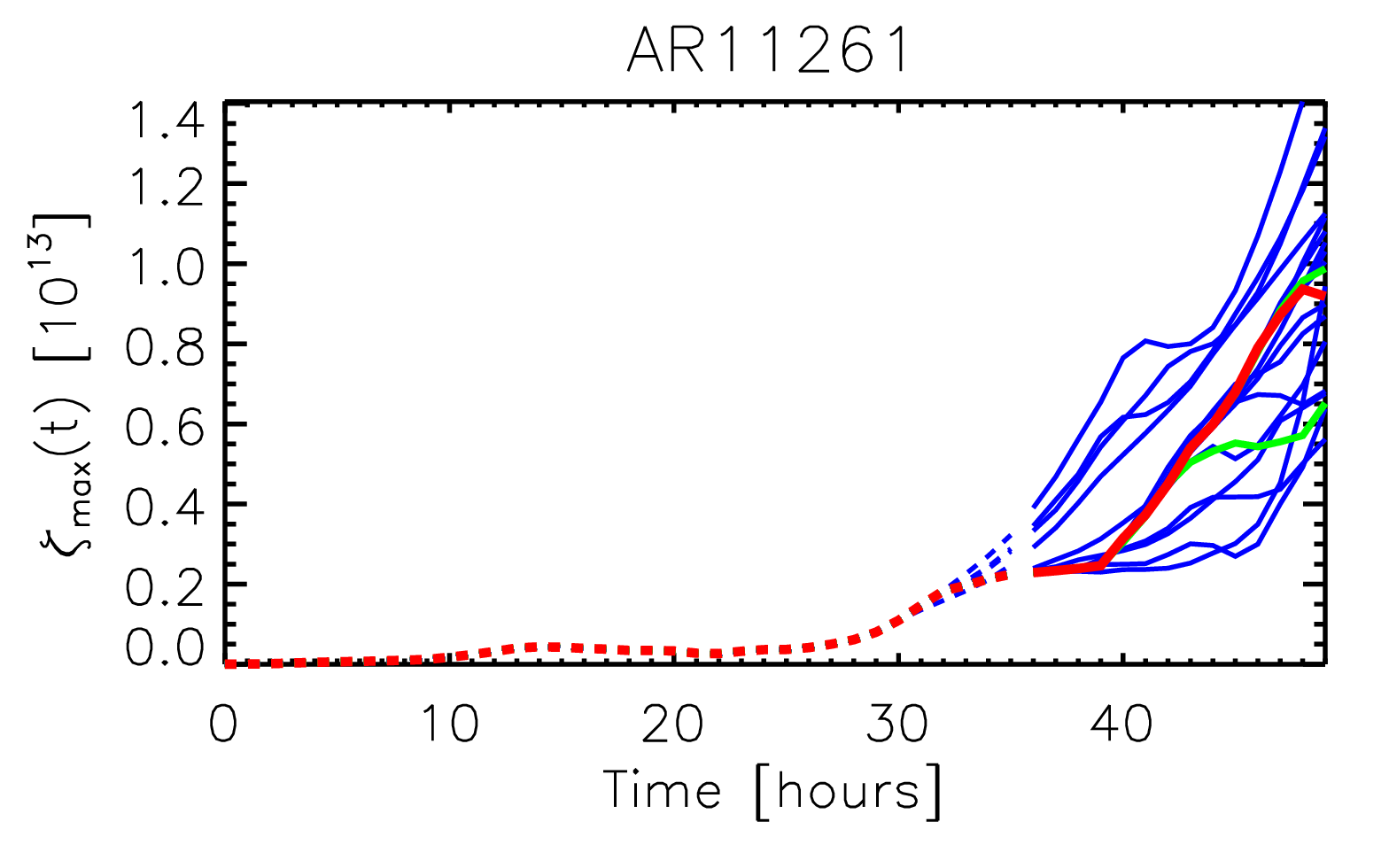}
\includegraphics[scale=0.28]{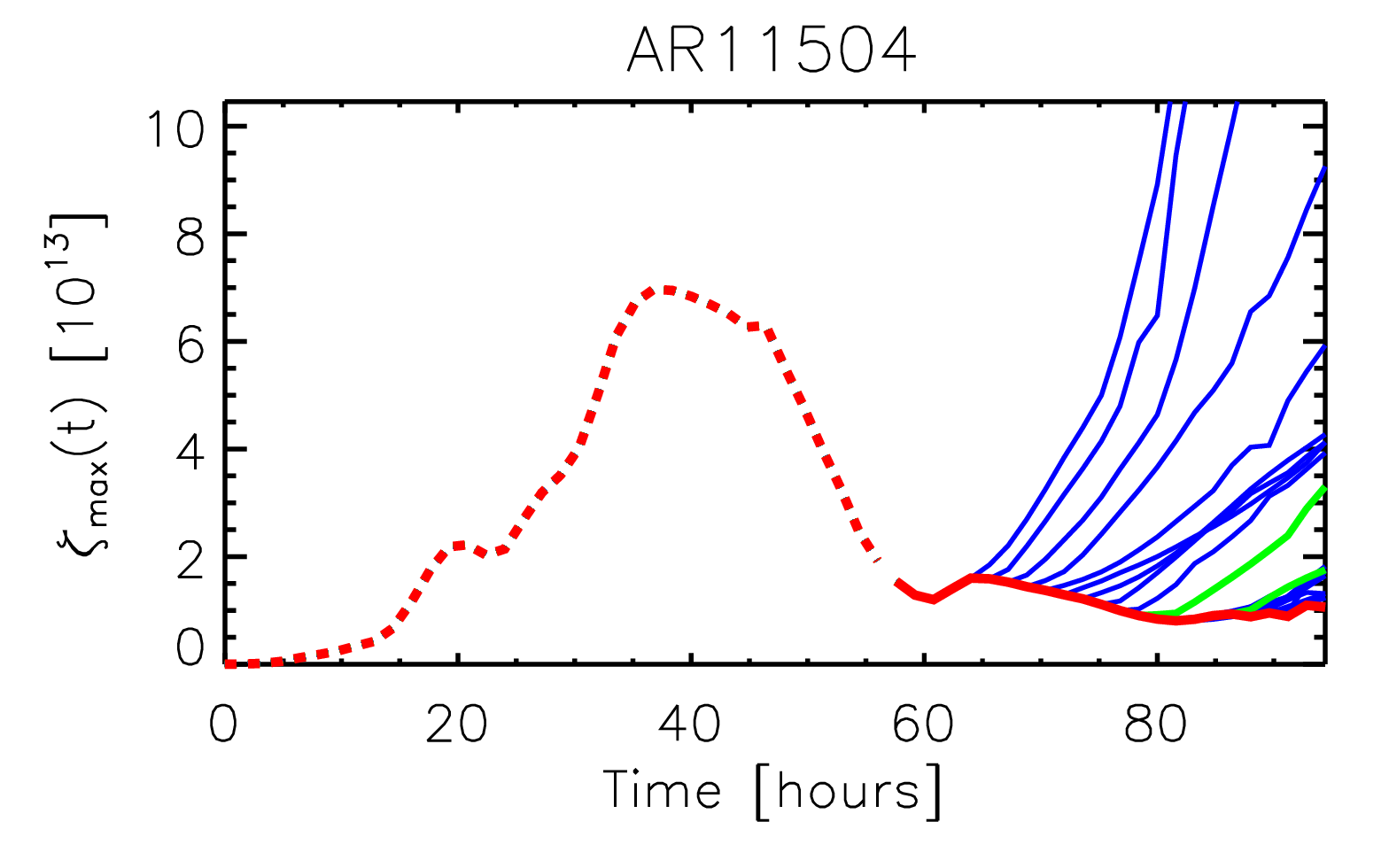}

\caption{The evolution of $\zeta_{max}\left(t\right)$ 
for the eruptive regions in our dataset (derived from non-normalised quantities).
The time $t_0$ when we switch from observed to projected magentograms is varied (blue curves).
The red curve represents the evolution for the reference simulations where $t_0=t_f$ and
the dashed red curve represents the evolution in the ramp-up phase.
The green curves show the evolution when $t_0=t_f-10\Delta t$ and $t_0=t_f-5\Delta t$.}
\label{taueruptive}
\end{figure}

For the purpose of this work, we are interested in studying
the evolution of our eruption metric $\zeta$, when we have replaced observed magnetograms with projected ones.
These numerical experiments are useful 
in understanding how much the evolution of our metric depends
on the long term evolution of the active region.
Fig.\ref{taueruptive} shows the time evolution of $\zeta_{max}\left(t\right)$ 
for the simulations concerning the five eruptive active regions.
Unlike Fig.\ref{critevol}, we do not normalise the value 
of $\omega\left(x,y,t\right)$, $\mu\left(x,y,t\right)$, and $\sigma\left(x,y,t\right)$,
in order to show how the evolution differs, when
varying the time at which we switch between observed and projected magnetograms.
In Fig.\ref{taueruptive} the red curve represents the simulation where $t_0=t_f$
and the green and blue curves correspond to simulations using projected magnetograms. 
We find that the evolution of $\zeta_{max}\left(t\right)$ can
differ significantly depending on the length of projection.
The closer the time $t_0$ is to $t_f$, the closer the evolution of $\zeta_{max}\left(t\right)$ is to the simulation where $t_0=t_f$,
i.e. the cases with the least amount of projection lead to the smallest differences compared to the full observational data case.
In most of the cases, when we introduce projected magnetograms, we obtain larger values of $\zeta_{max}\left(t\right)$, as our projection technique is equivalent to a persistent electric field in each magnetogram pixel. In contrast, we expect that
the magnetic field variations at the lower boundary are not always constant over an extended period of time.
For this reason, the evolution of $\zeta_{max}\left(t\right)$ shows its most significant departures when 
flux emergence occurs in the projected magnetograms,
as these events are assumed to persist over the full projection time.
AR11680 has the longest time series of observed magnetograms where
projected magnetograms are used only after around 90 observed magnetograms
($\sim6$ days).
However, the evolution of $\zeta_{max}\left(t\right)$ is the least scattered.
This emphasises the importance of having continuous long-lasting data sets,
where the persistence of information can be maintained in the coronal field.
In contrast, AR11561 and AR11504 have the shortest observed magnetogram sequence before projected magnetograms are used and the corresponding evolution of $\zeta_{max}\left(t\right)$ is highly scattered.
In Fig.\ref{taueruptive} the green curves show the simulations
associated with $t_0=t_f-5\Delta t$ and $t_0=t_f-10\Delta t$.
For AR11561, this is the $\zeta_{max}\left(t\right)$ evolution associated with the images in
the centre and right columns of Fig.\ref{comparisonB_AR11561}.
It is remarkable that in spite of the differences between the panels in Fig.\ref{comparisonB_AR11561},
the evolution of $\zeta_{max}\left(t\right)$ does
not significantly change when $t_0=t_f-10\Delta t$. This is true for most of the 
active regions with the exception of AR11437.
AR11437 shows observational signatures of 
an eruption about 10 hours before the end of the magnetogram series
which corresponds to a phase of increasing $\zeta_{max}\left(t\right)$.
In light of this, $\zeta_{max}\left(t\right)$ decreases after the eruption because the system has released energy leading to a decrease in the Lorentz force and the magnetic field complexity.
The simulation fails to describe this evolution when projected magnetograms are used as the lower boundary conditions.

\begin{figure}
\centering
\includegraphics[scale=0.28]{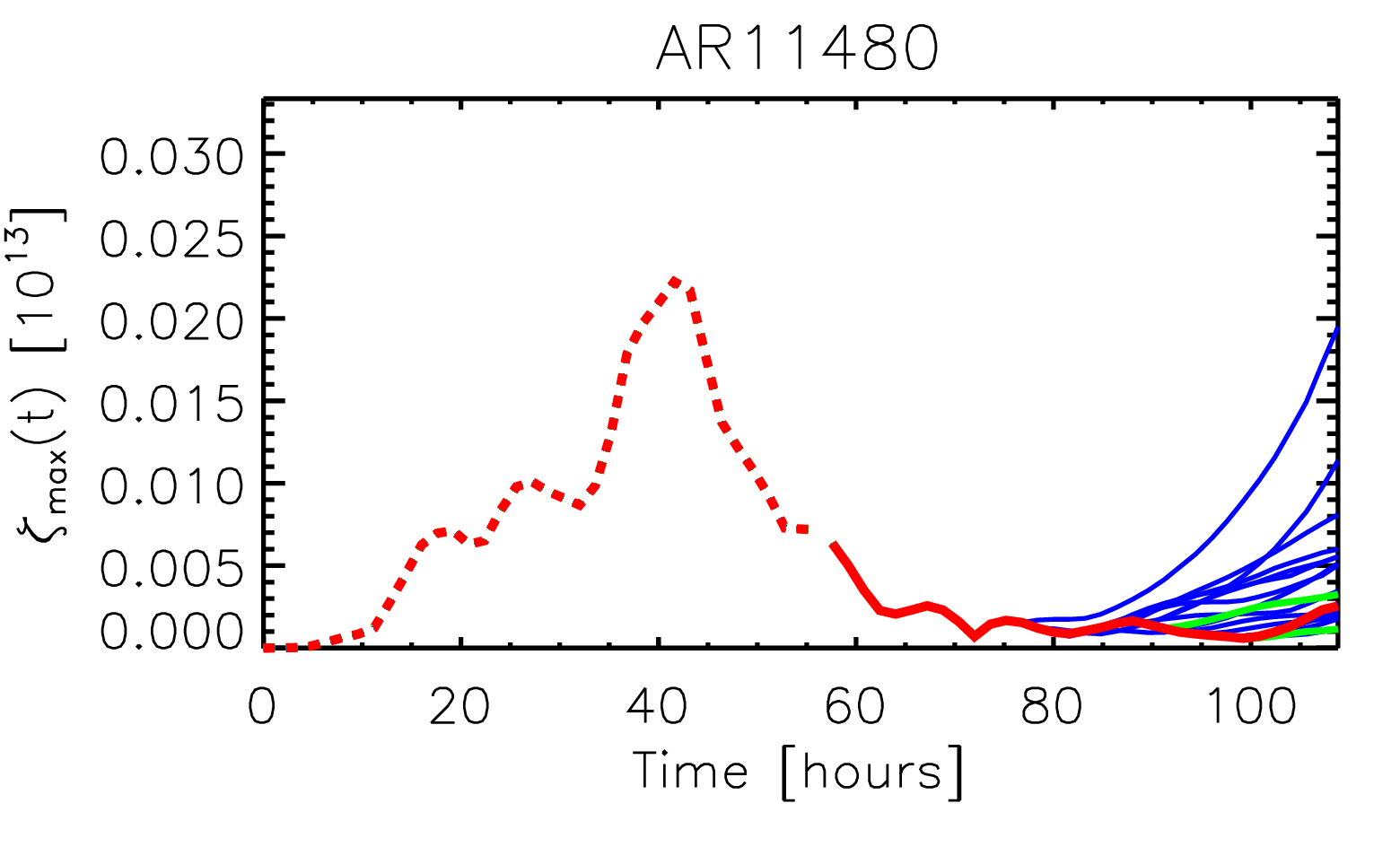}
\includegraphics[scale=0.28]{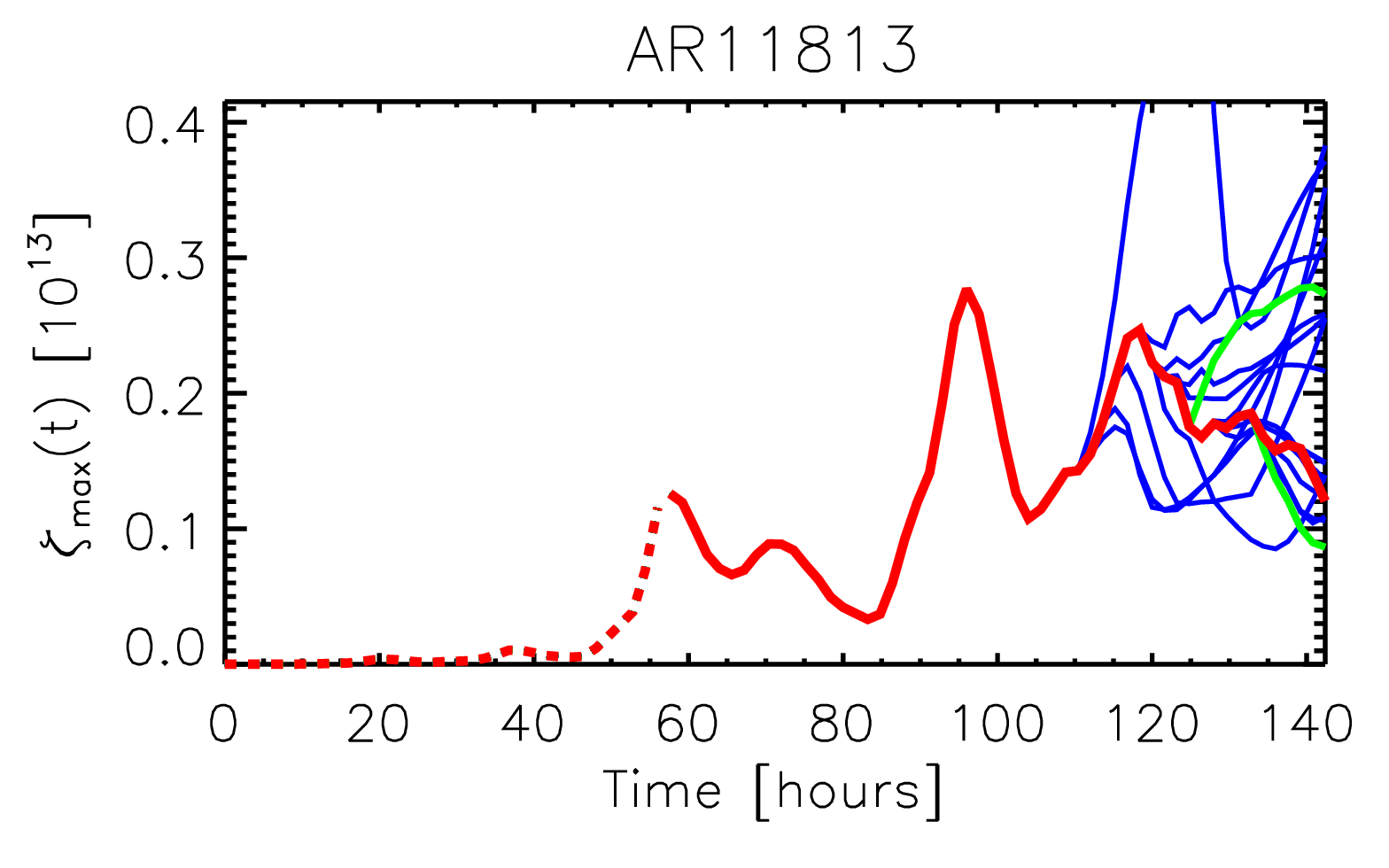}
\includegraphics[scale=0.28]{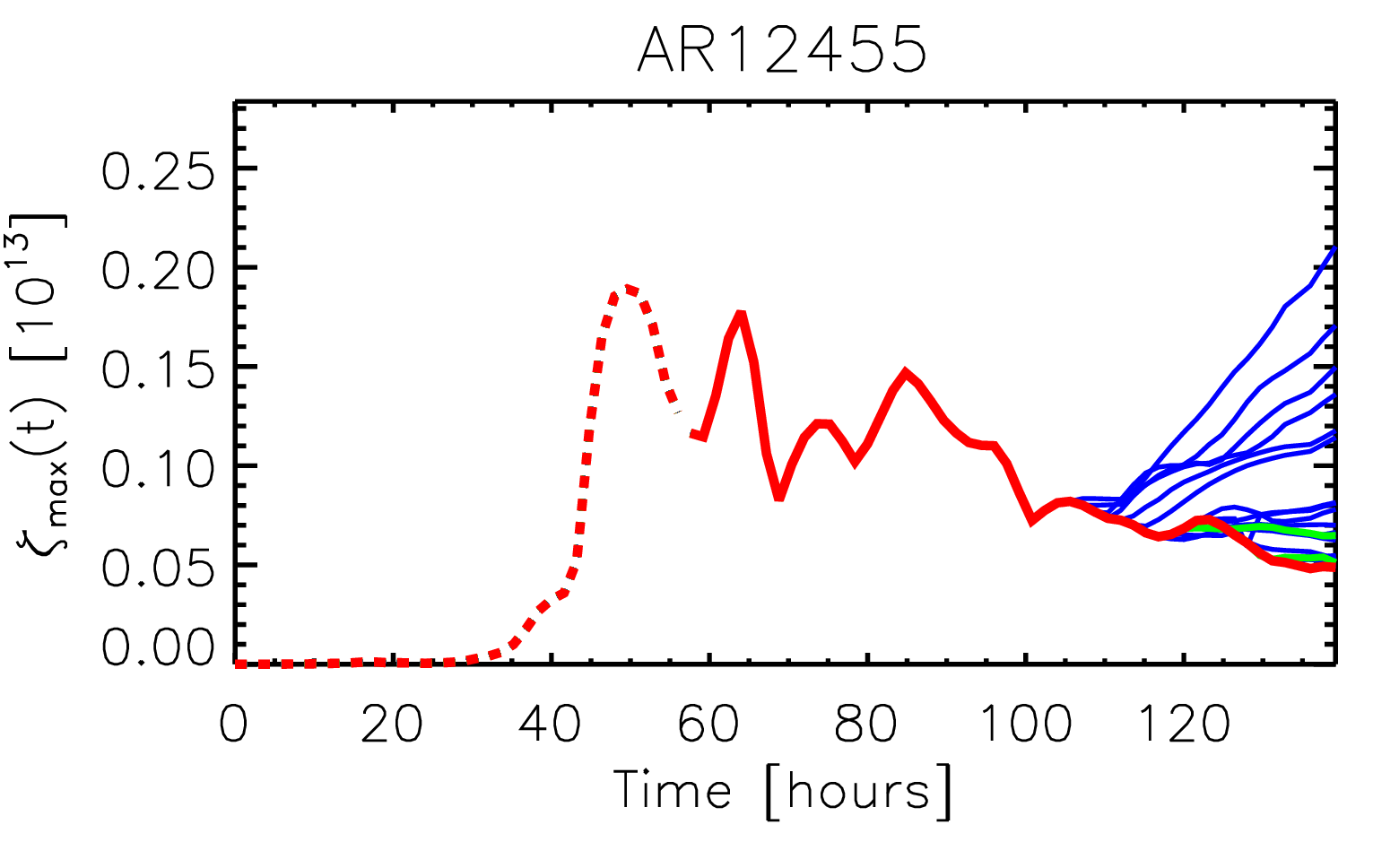}
\caption{The evolution of $\zeta_{max}\left(t\right)$ 
for the non-eruptive regions in our dataset (derived from non-normalised quantities).
The time $t_0$ when we switch from observed to projected magentograms is varied (blue curves).
The red curve represents the evolution for the reference simulations where $t_0=t_f$ and
the dashed red curve represents the evolution in the ramp-up phase.
The green curves show the evolution when $t_0=t_f-10\Delta t$ and $t_0=t_f-5\Delta t$.}\label{taunoneruptive}
\end{figure}
Similar conclusions can be drawn from Fig.\ref{taunoneruptive} which shows the same evolution of $\zeta_{max}\left(t\right)$ for the non-eruptive active regions.
In simulations which use projected magnetograms, the evolution of $\zeta_{max}\left(t\right)$ tends to deviate from the reference simulation following the slope at the time when projected magnetograms are introduced.
The evolution of $\zeta_{max}\left(t\right)$ can differ substantially when projected magnetograms are introduced, however the
simulation results diverge less from the reference simulation as $t_0$ approaches $t_f$.

\begin{figure}
\centering
\includegraphics[scale=0.18]{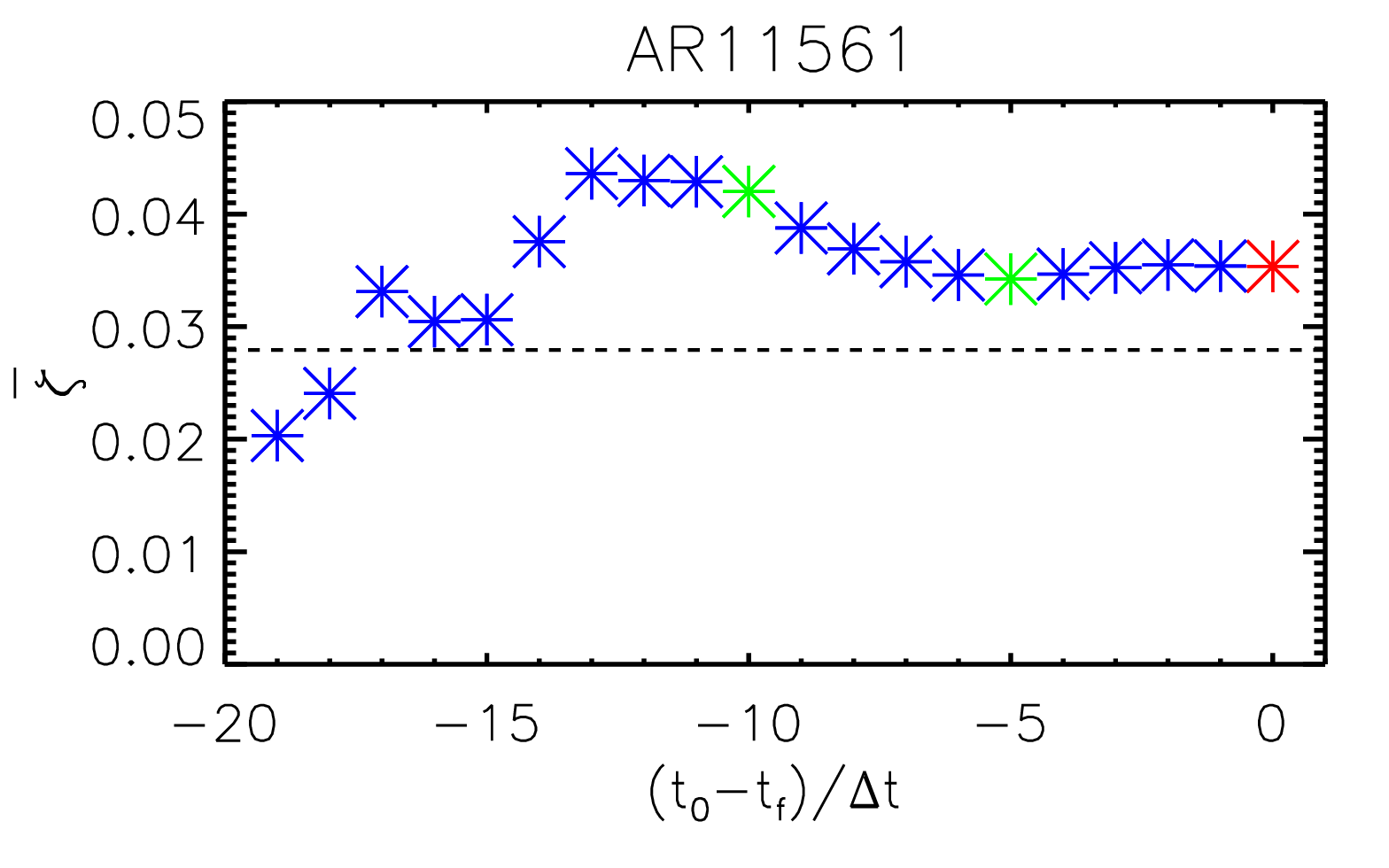}
\includegraphics[scale=0.18]{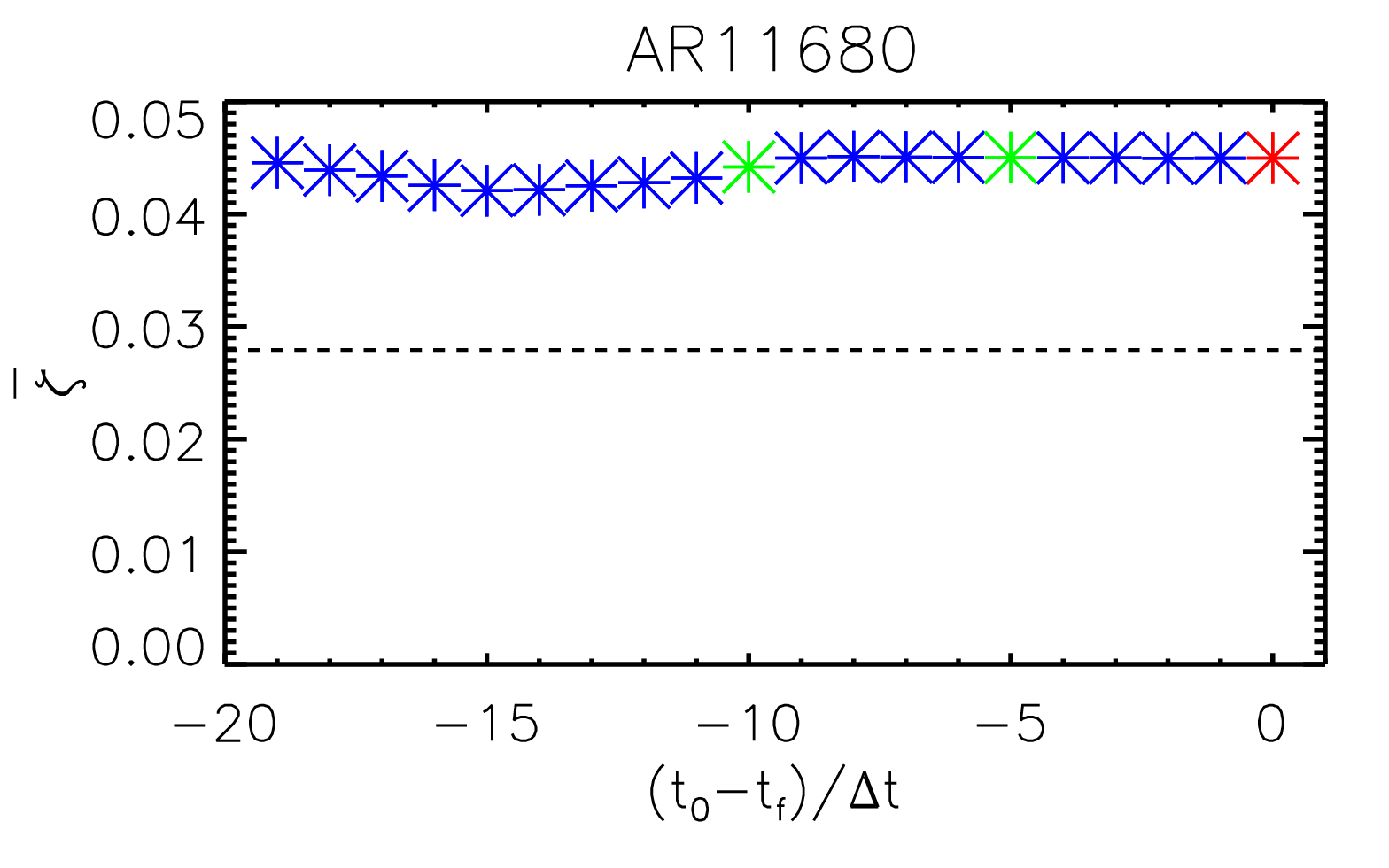}
\includegraphics[scale=0.18]{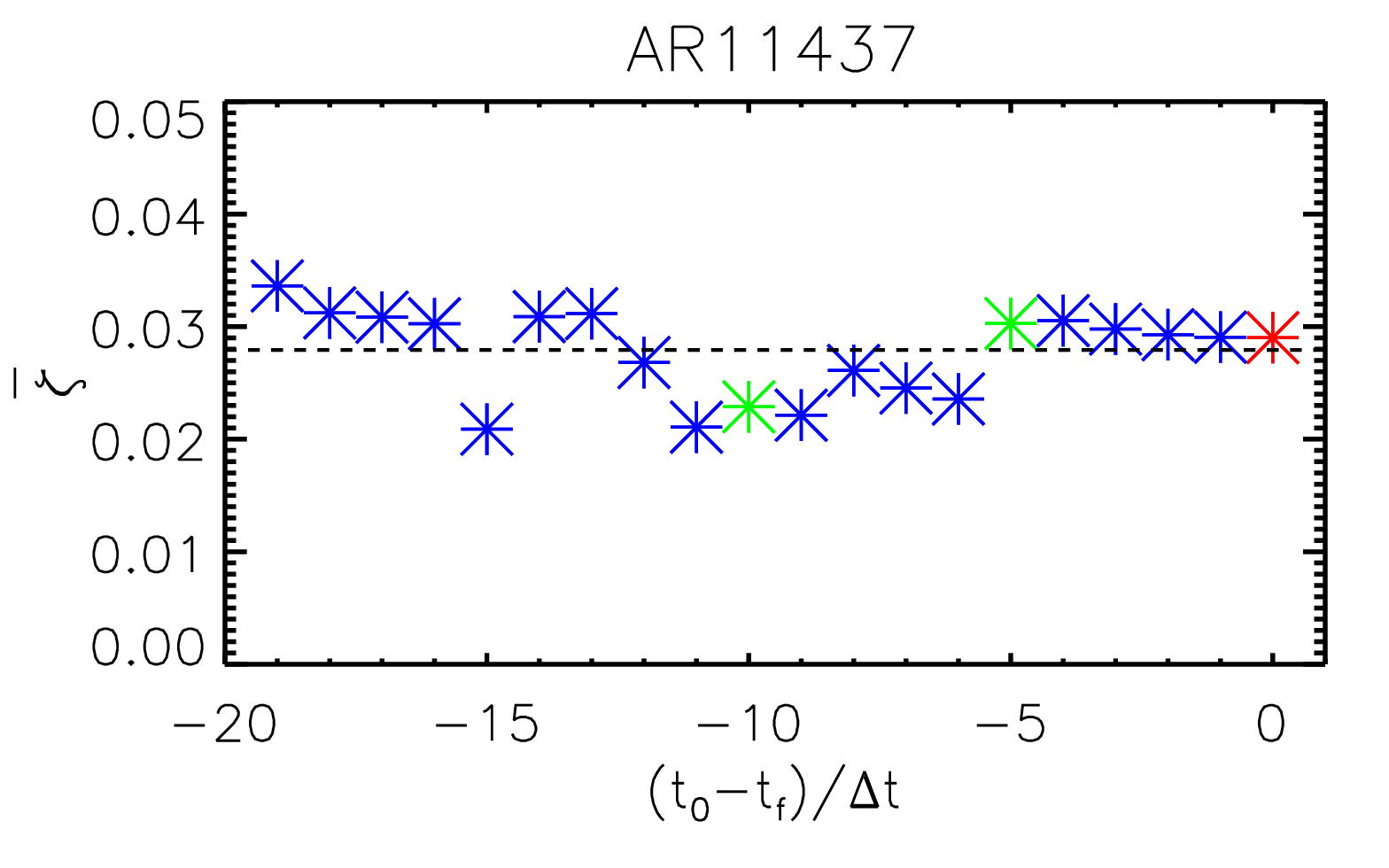}
\includegraphics[scale=0.18]{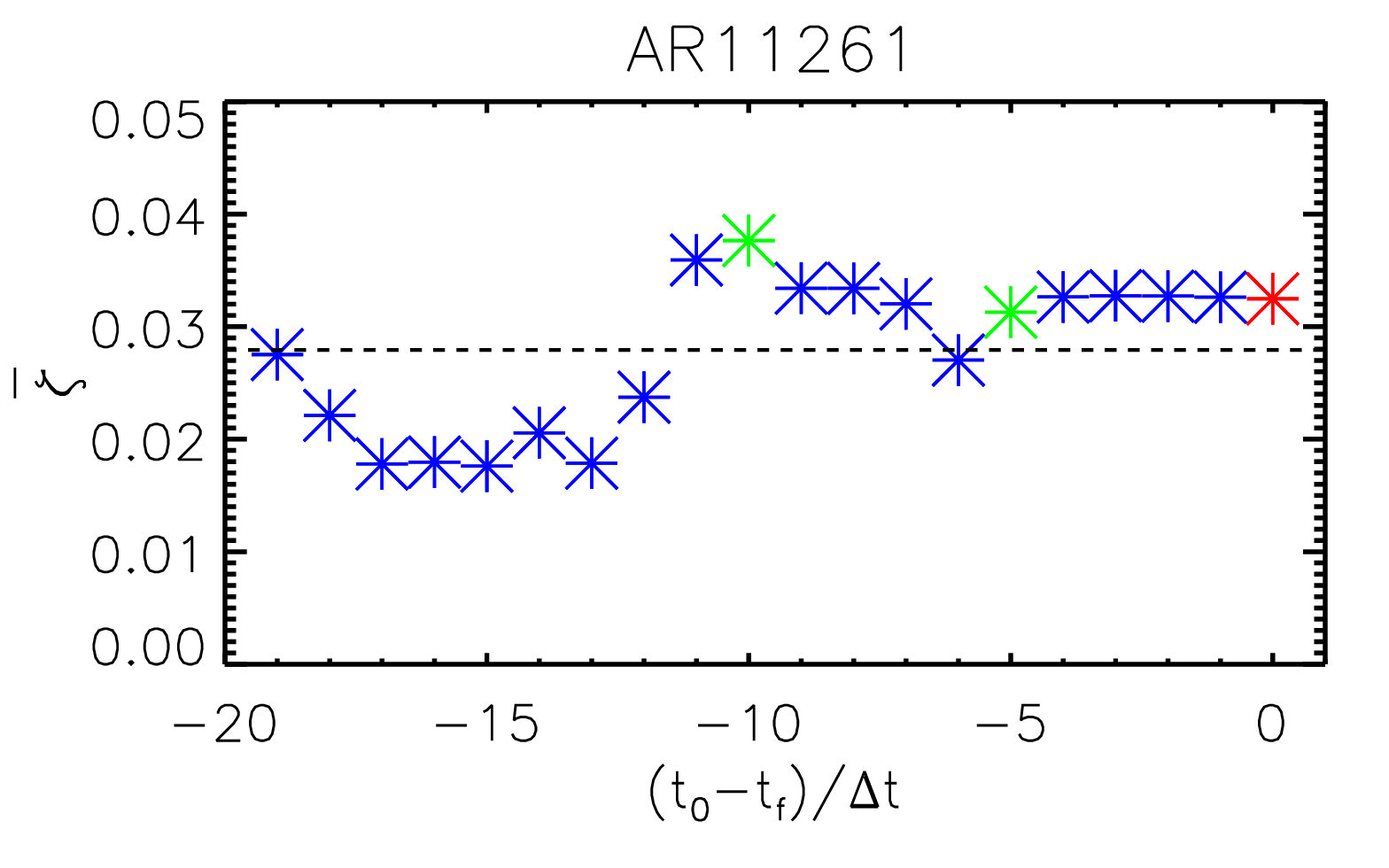}
\includegraphics[scale=0.18]{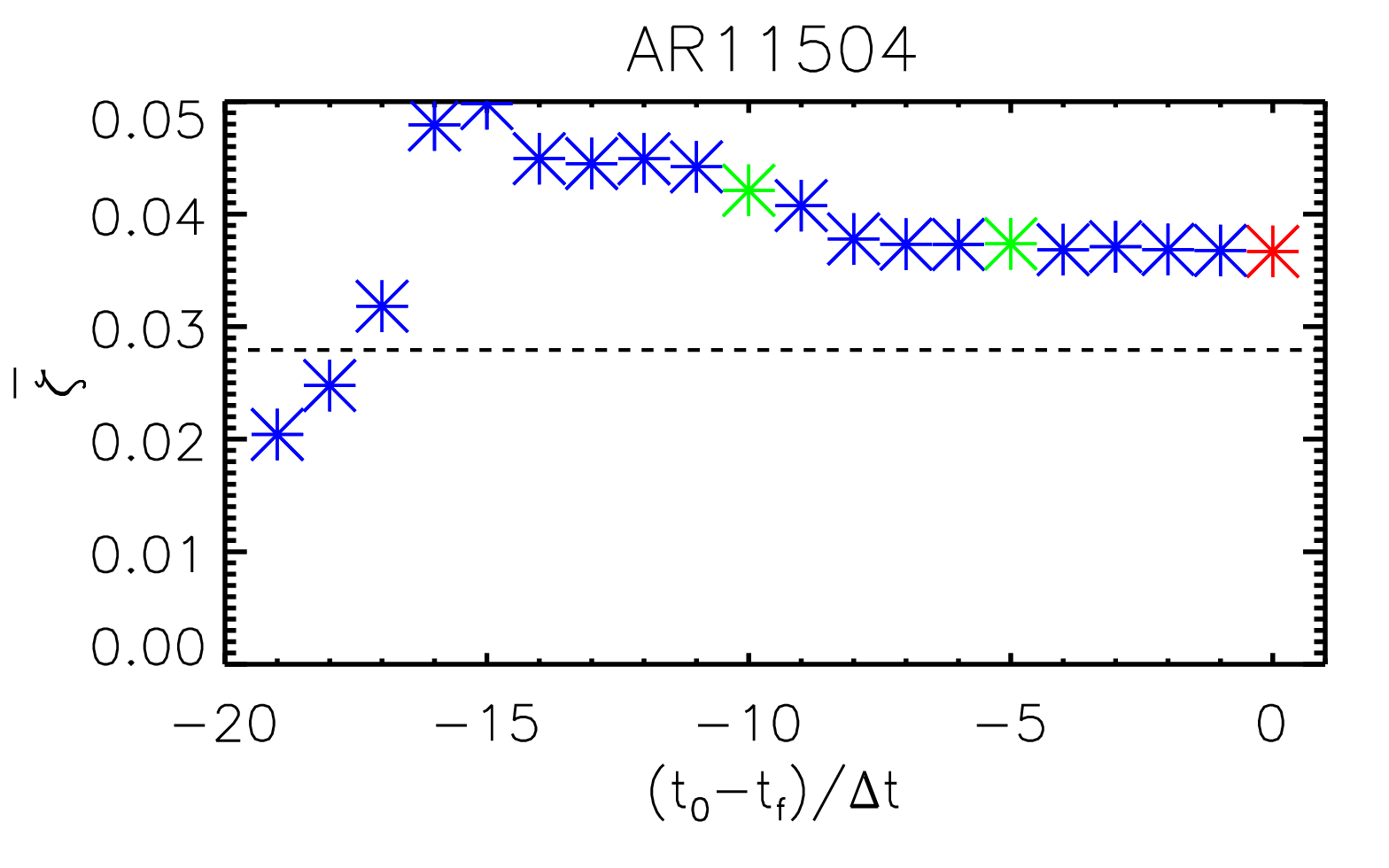}

\caption{The value of $\bar{\zeta}$ for each eruptive active region as a function of $t_0$, i.e. when we switch from observed to projected magentograms.
The dashed curve represents the value of $\bar{\zeta_{th}}=0.028$.
The red asterisk represents the simulation with $t_0=t_f$ and
the green asterisks represent the simulations when $t_0=t_f-5\Delta t$ and $t_0=t_f-10\Delta t$.}
\label{critprediction_eruptive}
\end{figure}

Another goal of applying a projected evolution 
is to identify, in advance, active regions that will erupt.
In Sec.\ref{criteria}, we concluded that
the parameter $\bar{\zeta}$ (the time average of $\zeta_{max}\left(t\right)$)
best discriminates eruptive from non-eruptive active regions.
Therefore, we compare the value of $\bar{\zeta}$
obtained for the simulations using only observed magnetograms 
with the ones using projected magnetograms over different projection timescales.
Fig.\ref{critprediction_eruptive} shows the value of 
$\bar{\zeta}$ for each simulation using projected magnetograms (blue asterisks)
in comparison with the simulation with only observed magnetograms (red asterisk),
as a function of $t_0$ for the eruptive active regions.
We use green asterisks to signify $\bar{\zeta}$ for the simulations with
$t_0=t_f-5\Delta t$ and $t_0=t_f-10\Delta t$.
We find that in all active regions
where we use projected magnetograms, the simulations converge to the true value of $\bar{\zeta}$,
as $t_0$ approaches $t_f$.
It is clear that the value of $\bar{\zeta}$ for most of the active regions is accurately reproduced by the predictive simulations when $t_0\ge t_f-10\Delta t$,
whereas for AR11261 and AR11437 it happens only when $t_0\ge t_f-5\Delta t$.

If we compare the value of $\bar{\zeta}$ to the
threshold value $\bar{\zeta_{th}}$,
we find that for the majority of $t_0$ $\bar{\zeta}$ remains larger than $\bar{\zeta_{th}}$,
indicating the possible occurrence of an eruption.
For some cases the value of $\bar{\zeta}$ oscillates about the threshold, although
it always exhibits significant time periods where it is above the threshold.
For active regions AR11561, AR11261, and AR11504,
the predictions of $\bar{\zeta}$ result in higher $\bar{\zeta}$
compared to the simulation where $t_0=t_f$.
This occurs when flux emergence is ongoing in the active region at the time we switch from observed to projected magnetograms.
This is due to the simple projection technique applied which leads to a continuous increase of magnetic flux and magnetic stress.
When magnetic flux is not emerging, the values of $\bar{\zeta}$ can be predicted more accurately in advance.
To improve the accuracy of this approach significantly
a projection technique  
that mitigates the effect of magnetic flux emergence over long periods of time must be developed.

\begin{figure}
\centering
\includegraphics[scale=0.18]{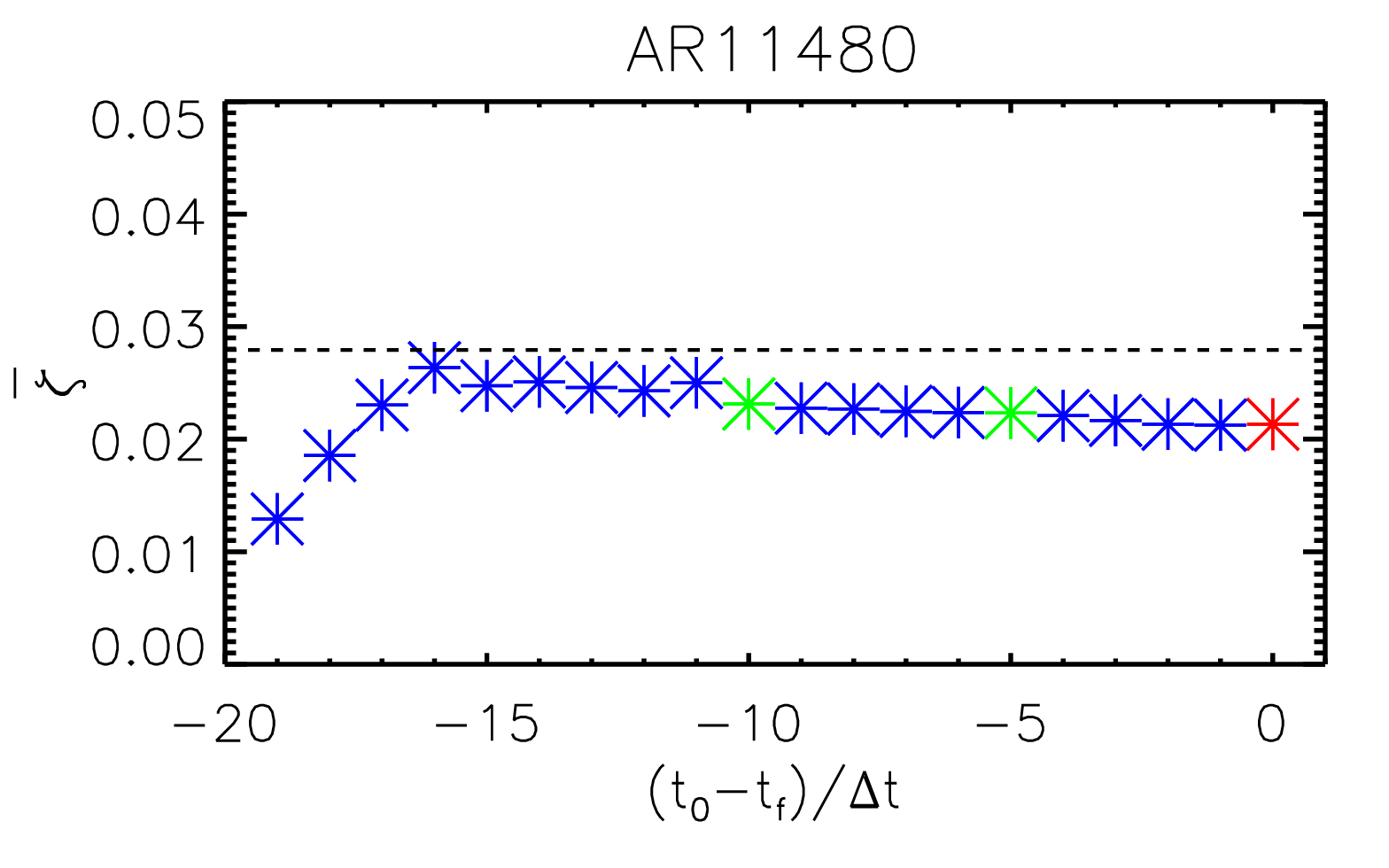}
\includegraphics[scale=0.18]{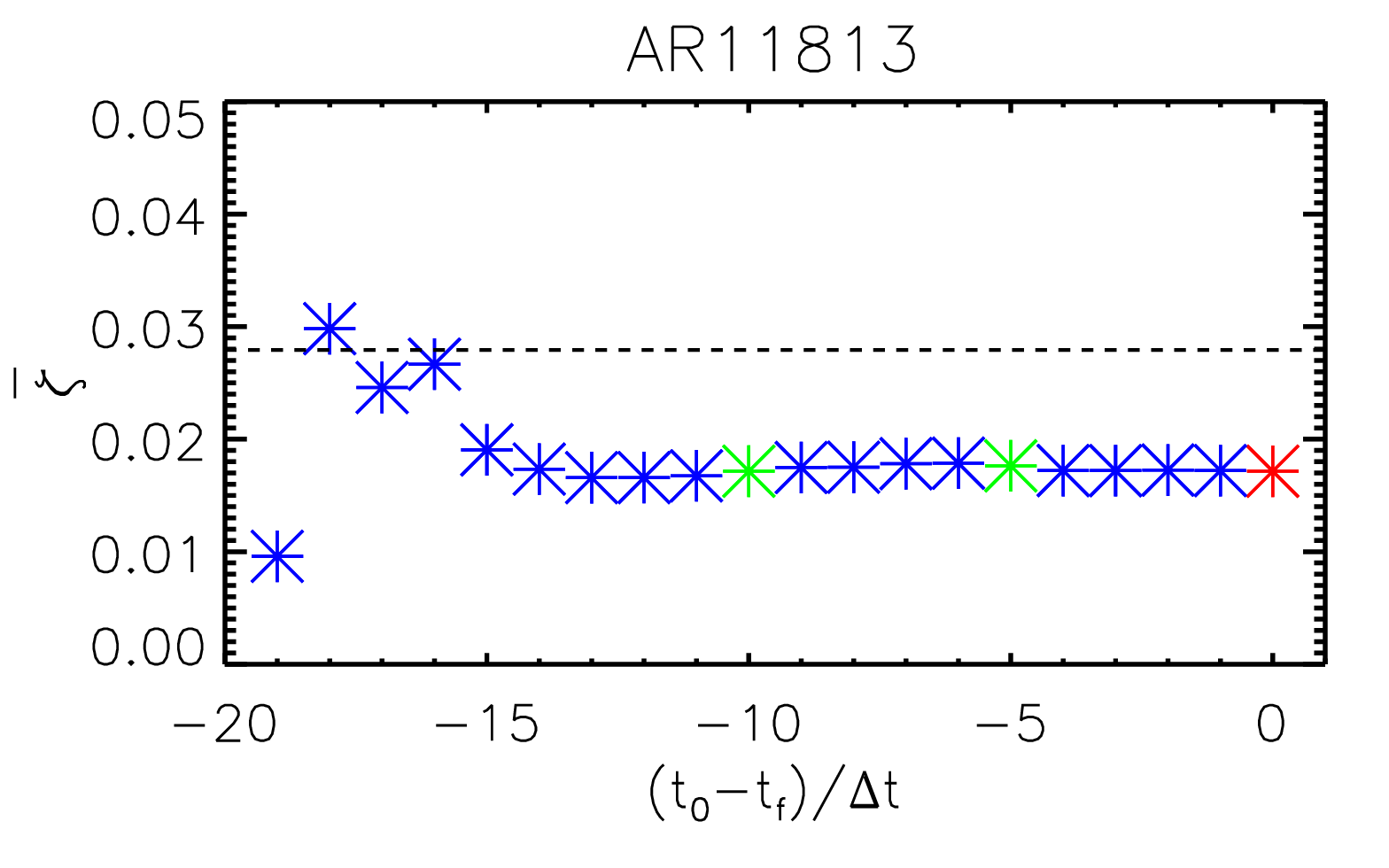}
\includegraphics[scale=0.18]{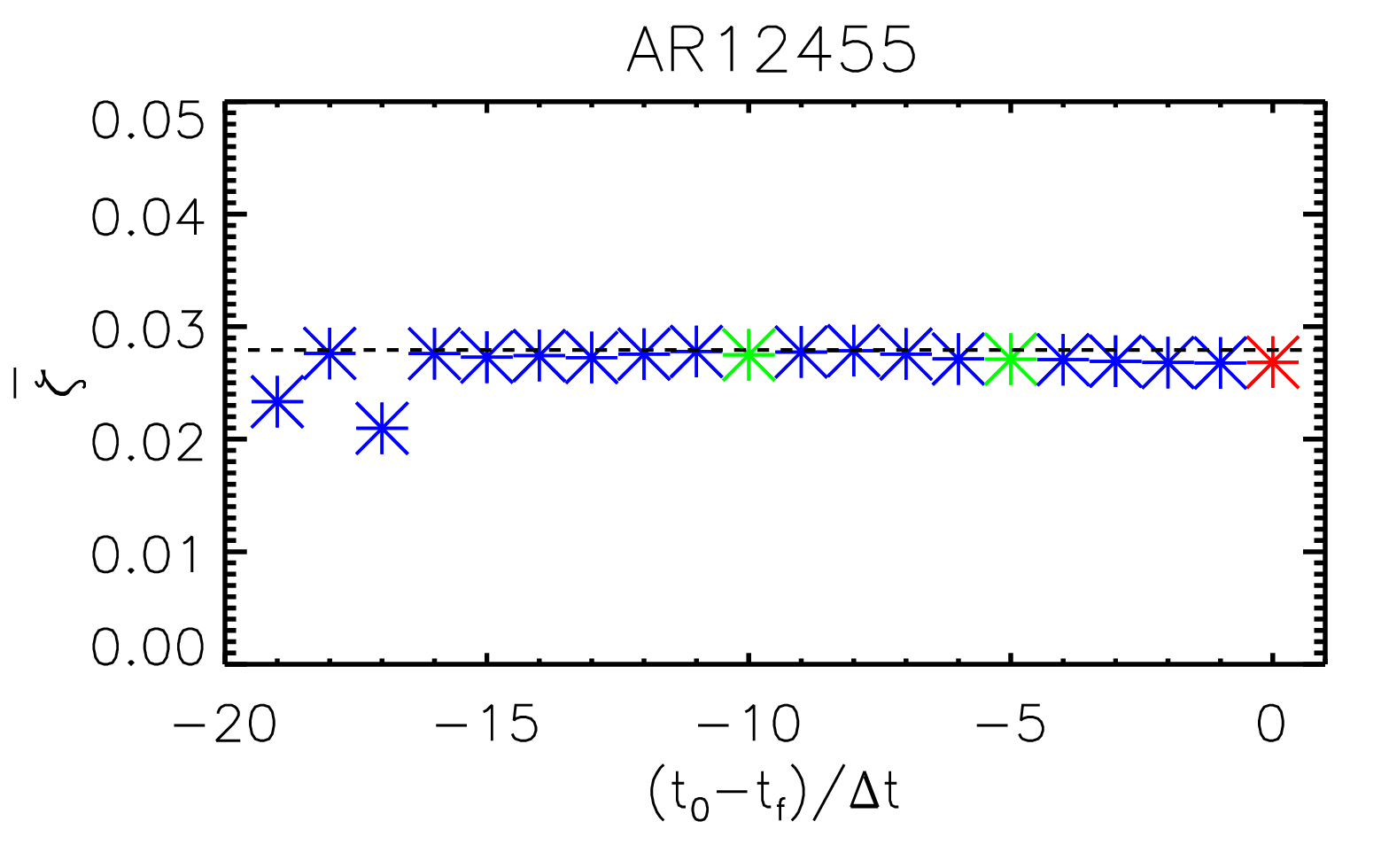}
\caption{The value of $\bar{\zeta}$ for each non-eruptive active region as a function of $t_0$, i.e. when we switch from observed to projected magentograms.
The dashed curve represents the value of $\bar{\zeta_{th}}=0.028$.
The red asterisk represents the simulation with $t_0=t_f$ and
the green asterisks represent the simulations when $t_0=t_f-5\Delta t$ and $t_0=t_f-10\Delta t$.}
\label{critprediction_noneruptive}
\end{figure}

Fig.\ref{critprediction_noneruptive} shows the same plot for the non-eruptive active regions 
confirming that the final value of $\bar{\zeta}$ can be estimated several time steps before the final magnetogram.
In general, the whole set of predictive simulations show a behaviour 
consistent with the simulation using only observed magnetograms,
where the value of $\bar{\zeta}$ remains lower than the threshold value.
Again we find some values of $t_0$ where the
predicted value of $\bar{\zeta}$
is significantly different from the simulation at $t_0=t_f$,
but they are rather isolated or occur over the longest predictive time scales,
where the use of observational information is limited.

Fig.\ref{critprediction_eruptive} and Fig.\ref{critprediction_noneruptive} show that the parameter $\bar{\zeta}$ can consistently identify eruptive from non-eruptive active regions.
With some limitations, this is also true when we replace observed magnetograms with projected ones.
This result emphasises the potential of the technique in
(i) selecting which active regions are most or least likely to erupt
or (ii) comparing the likelihood of eruptions between two active regions.
In order to further assess the robustness of this approach, in the next section we investigate the role of additional random noise in simulations during the time frame of the projected magnetograms.

\subsection{Projections with random noise component}

To test the robustness of this modelling technique, we run additional simulations
where projected magnetograms are perturbed with a component of random noise.
The purpose of these numerical experiments is to
investigate how the predicted value of $\bar{\zeta}$ is affected by random perturbations.

We test the effect of random noise on two active regions from our set, AR11561 (eruptive) and AR11813 (non-eruptive).
As described in Eq.\ref{extrapolateA} the quantity that drives the variation of the lower boundary is the electric field. 
First, we compute the two components of the surface electric field,
$E_x\left(x,y\right)$ and $E_y\left(x,y\right)$ (i.e. right-hand-side in Eq.\ref{extrapolateA}) from the final two observed magnetograms. In the previous simulations presented in Sect.\ref{predictionresults}, $E_x\left(x,y\right)$ and $E_y\left(x,y\right)$
were kept constant in time over the projected evolution. In contrast,
to introduce a randomly varying electric field, we next compute the mean value of $E_x\left(x,y\right)$ and $E_y\left(x,y\right)$ 
and the standard deviation $\sigma_{Ex}$ and $\sigma_{Ey}$ of the electric field.
These values are computed over the whole computational domain, where we note that
the strong field regions only occupy a small subset of the domain.
Finally, a random noise component is added to 
$E_x\left(x,y\right)$ and $E_y\left(x,y\right)$ at each pixel where the random 
component is varied at the magnetogram acquisition cadence (i.e. 96 or 60 minutes).
It should be noted that the values of $\sigma_{Ex}$ and $\sigma_{Ey}$
are approximately two orders of magnitude smaller than the electric field values in the active region centre. This is a consequence of computing these values over the full domain, where the strong field regions are localised at the center of the domain.
In the simulations with noise, the noise component varies in time around the values of $E_x\left(x,y\right)$ and $E_y\left(x,y\right)$, thus its contribution averaged over time is asymptotically zero.
The purpose of this exercise is therefore to test the robustness of the metric
against localised fluctuations of the electric field and its temporal variations.
We run three simulations where the amplitude of the noise component 
is $E^{64}_{noise}=64\sigma_{E}$, $E^{256}_{noise}=256\sigma_{E}$, and $E^{512}_{noise}=512\sigma_{E}$
(where we replace $\sigma_{Ex}$ and $\sigma_{Ey}$ with $\sigma_{E}$ for simplicity of notation). Large values of the noise relative to the standard deviation are required as when the standard deviation is computed over the full domain it is very small.

\begin{figure}
\centering
\includegraphics[scale=0.23]{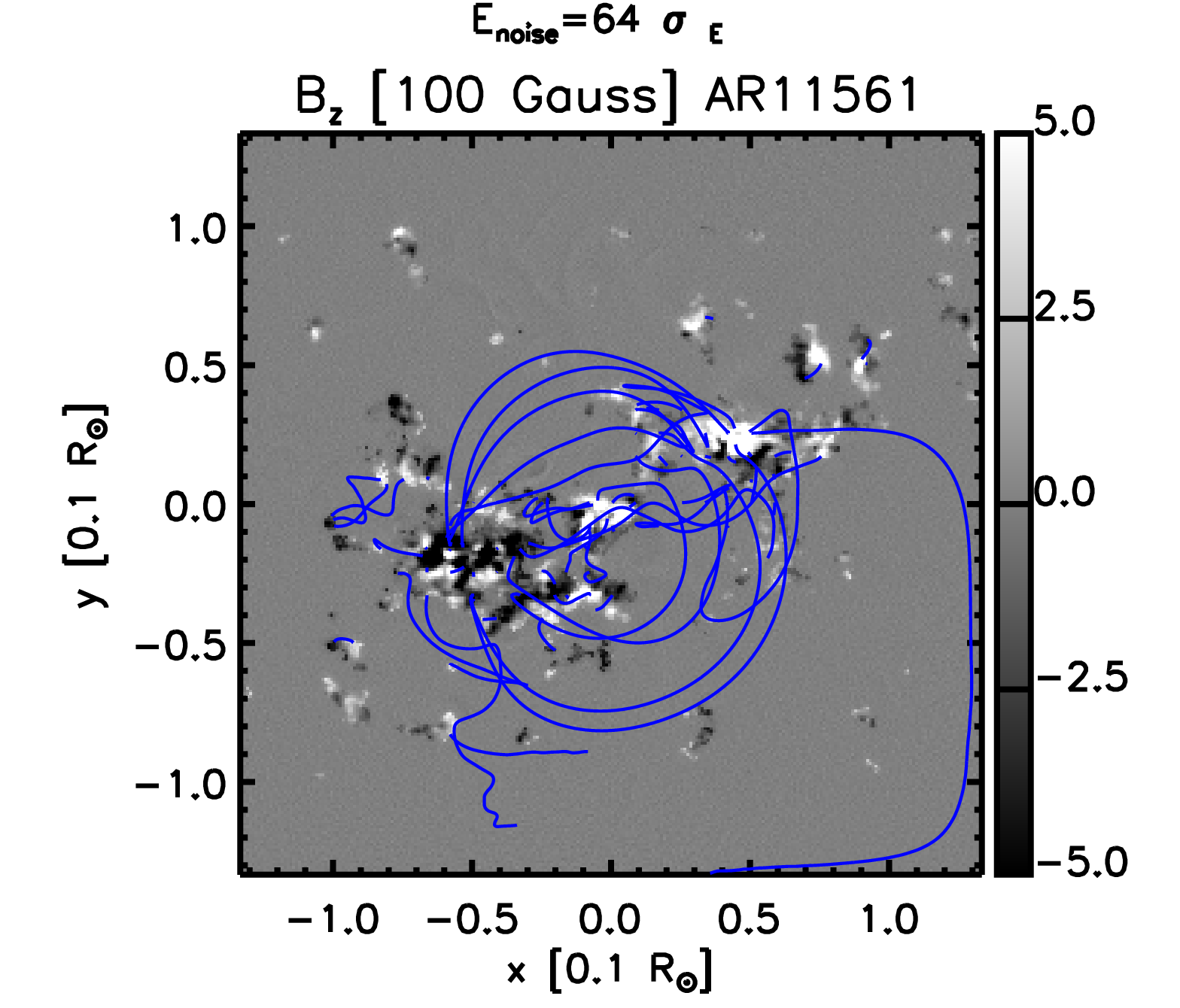}
\includegraphics[scale=0.23]{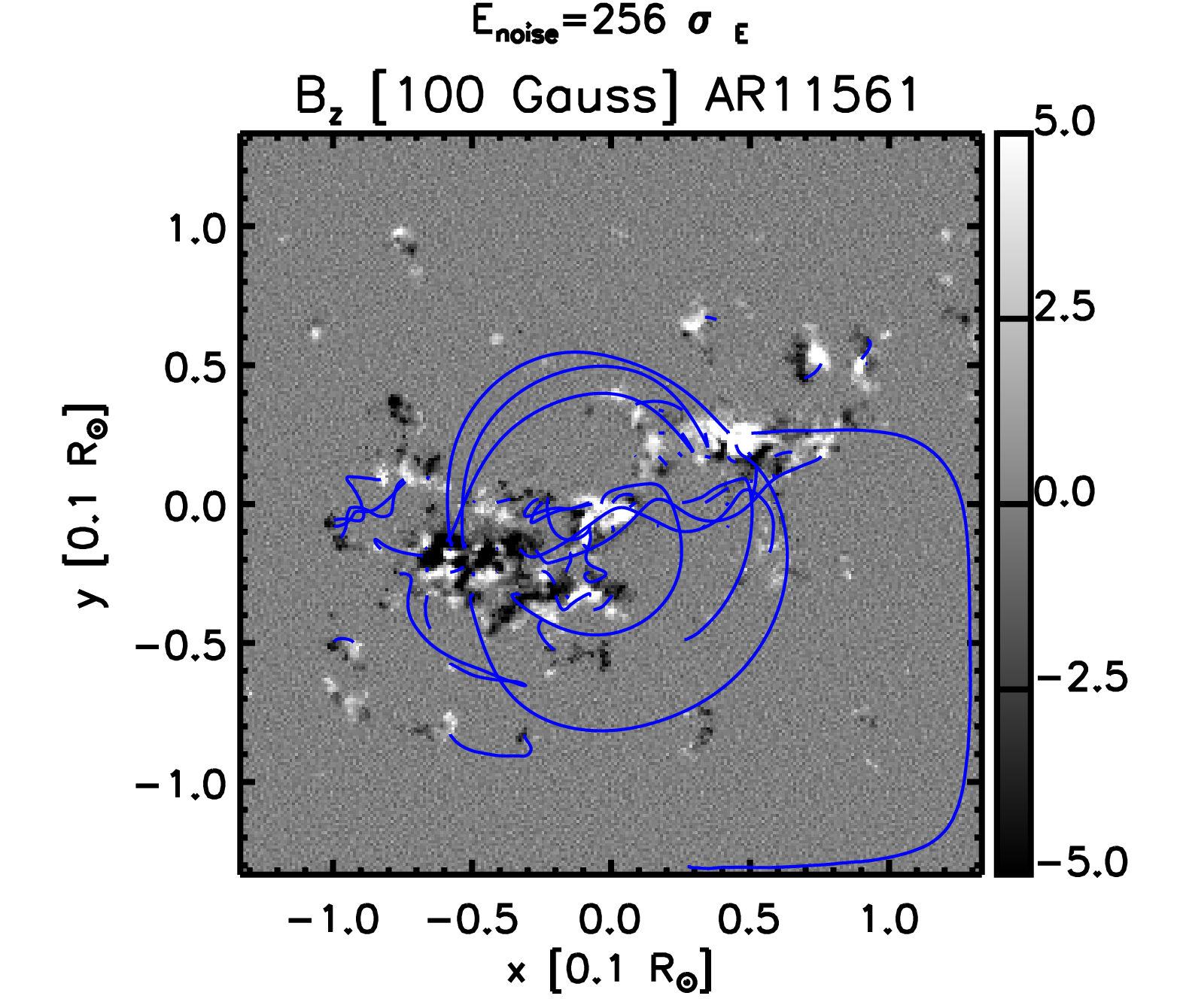}
\includegraphics[scale=0.23]{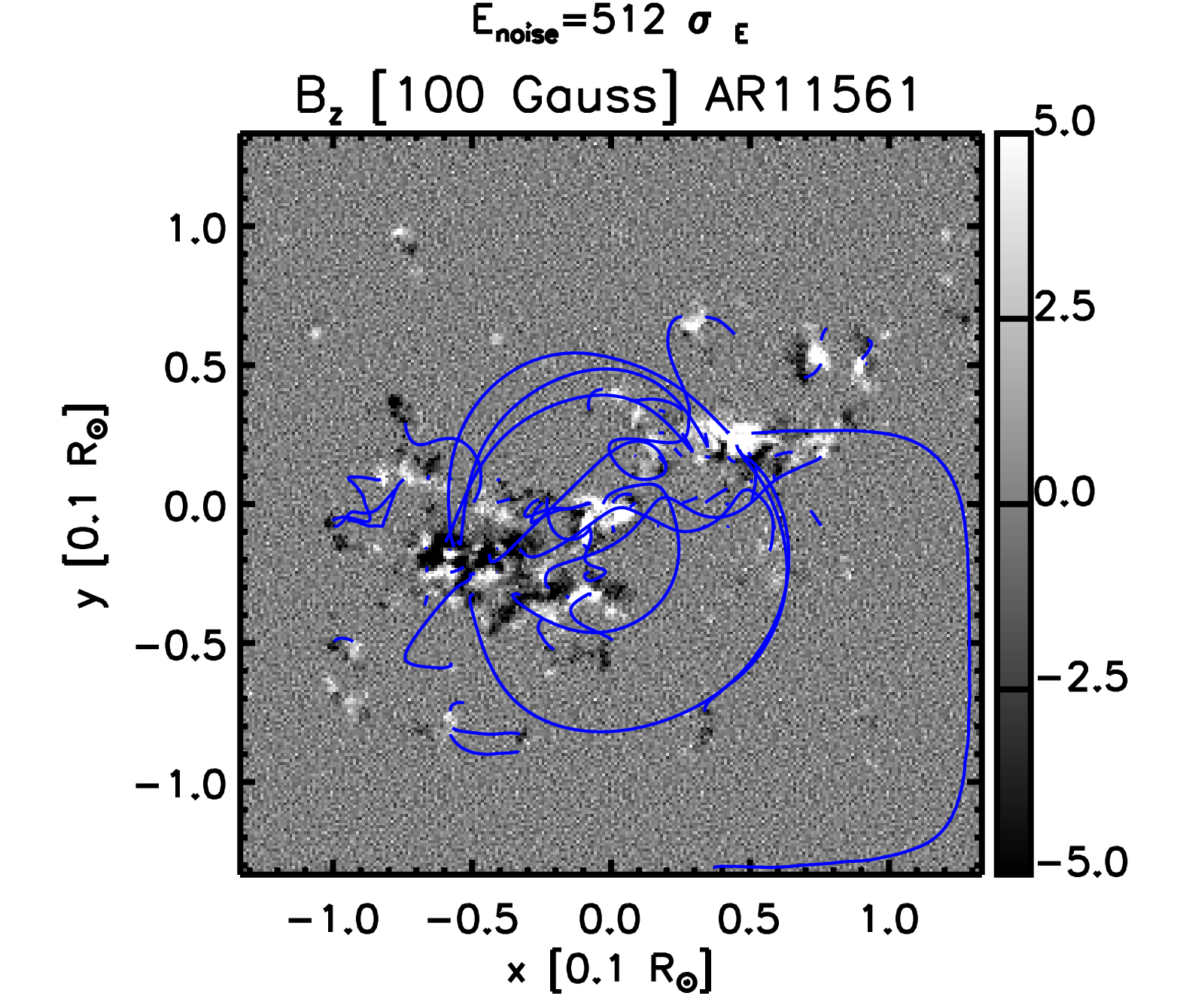}
\caption{Maps of the photospheric $B_z$ at $t=t_f$ with 
sample magnetic field curves overplotted for the three simulations
of AR11561 with $t_0=t_f-10\Delta t$, where we apply a random noise component in the projected magnetograms with values of $E^{64}_{noise}=64\sigma_{E}$ (left),
$E^{256}_{noise}=256\sigma_{E}$ (centre),
and $E^{512}_{noise}=512\sigma_{E}$ (right), respectively.}
\label{noise_AR11561}
\end{figure}

We present the results for the simulations with 
$t_0=t_f-5\Delta t$ and $t_0=t_f-10\Delta t$ for both of the active regions.
Fig.\ref{noise_AR11561} shows the final distribution of $B_z$ for the three simulations with
$E^{64}_{noise}=64\sigma_{E}$, $E^{256}_{noise}=256\sigma_{E}$, and $E^{512}_{noise}=512\sigma_{E}$
for $t_0=t_f-10\Delta t$ for AR11561.
The true observed magnetograms can be seen in Fig.\ref{bzeruptive} for comparison.
We find that most magnetic structures are not significantly affected by the noise.
The noise only becomes visible at the single pixel level when $E_{noise}=E^{512}_{noise}$.
We find similar results for AR11813 (not shown here).

\begin{figure}
\centering
\includegraphics[scale=0.30]{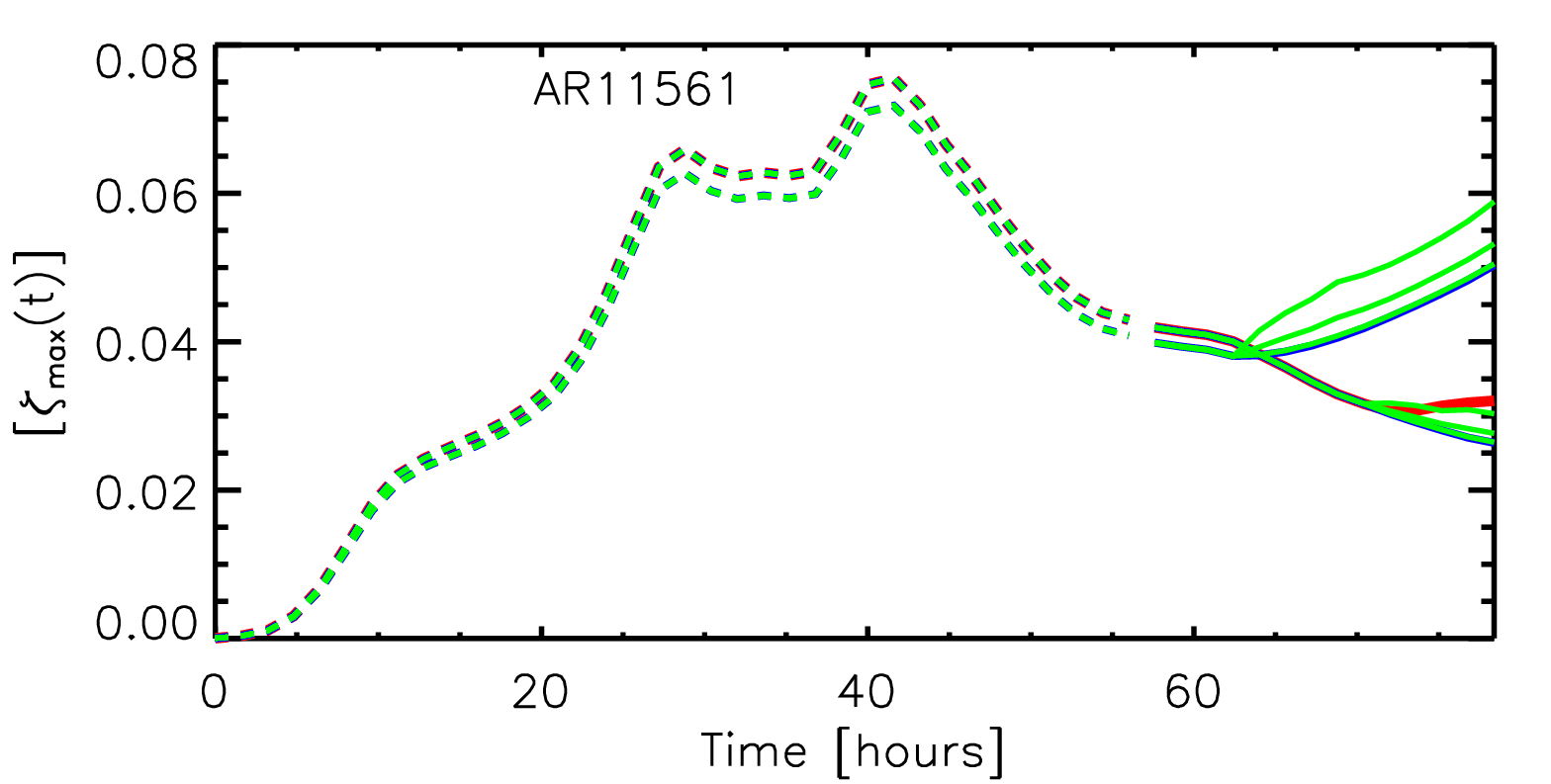}
\includegraphics[scale=0.30]{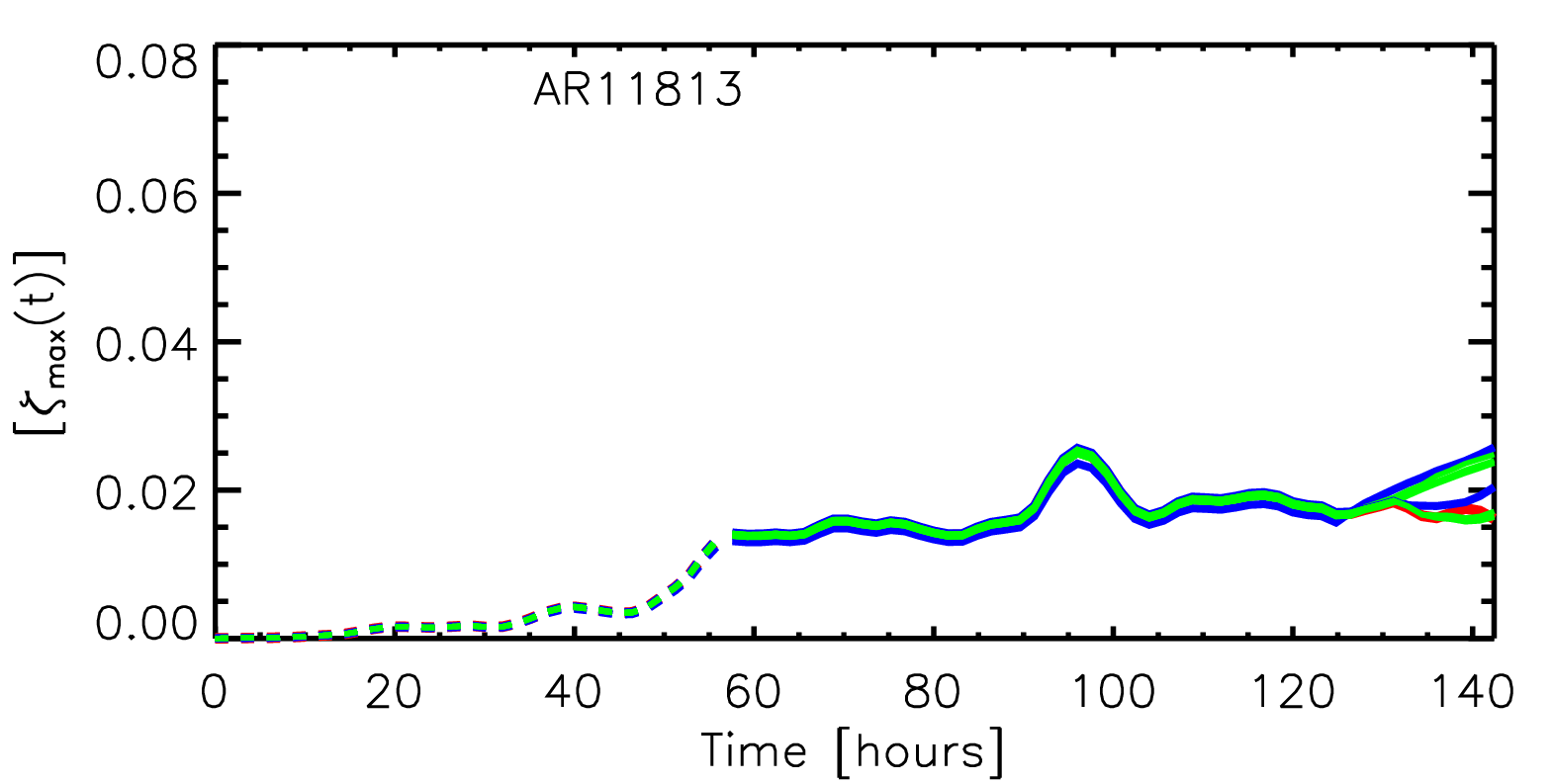}
\caption{Evolution of $\zeta_{max}\left(t\right)$
for AR11561 (left panel) and AR11813 (right panel).
The red curves represents the simulation where $t_0=t_f$,
the blue curves represent the simulations using projected magnetograms with $t_0=t_f-5\Delta t$ and $t_0=t_f-10\Delta t$,
and the green curves represent the simulations using the same values of $t_0$ with an added a noise component
of $E^{64}_{noise}=64\sigma_{E}$, $E^{256}_{noise}=256\sigma_{E}$, and $E^{512}_{noise}=512\sigma_{E}$.}
\label{noisetauevol}
\end{figure}

\begin{figure}
\centering
\includegraphics[scale=0.30]{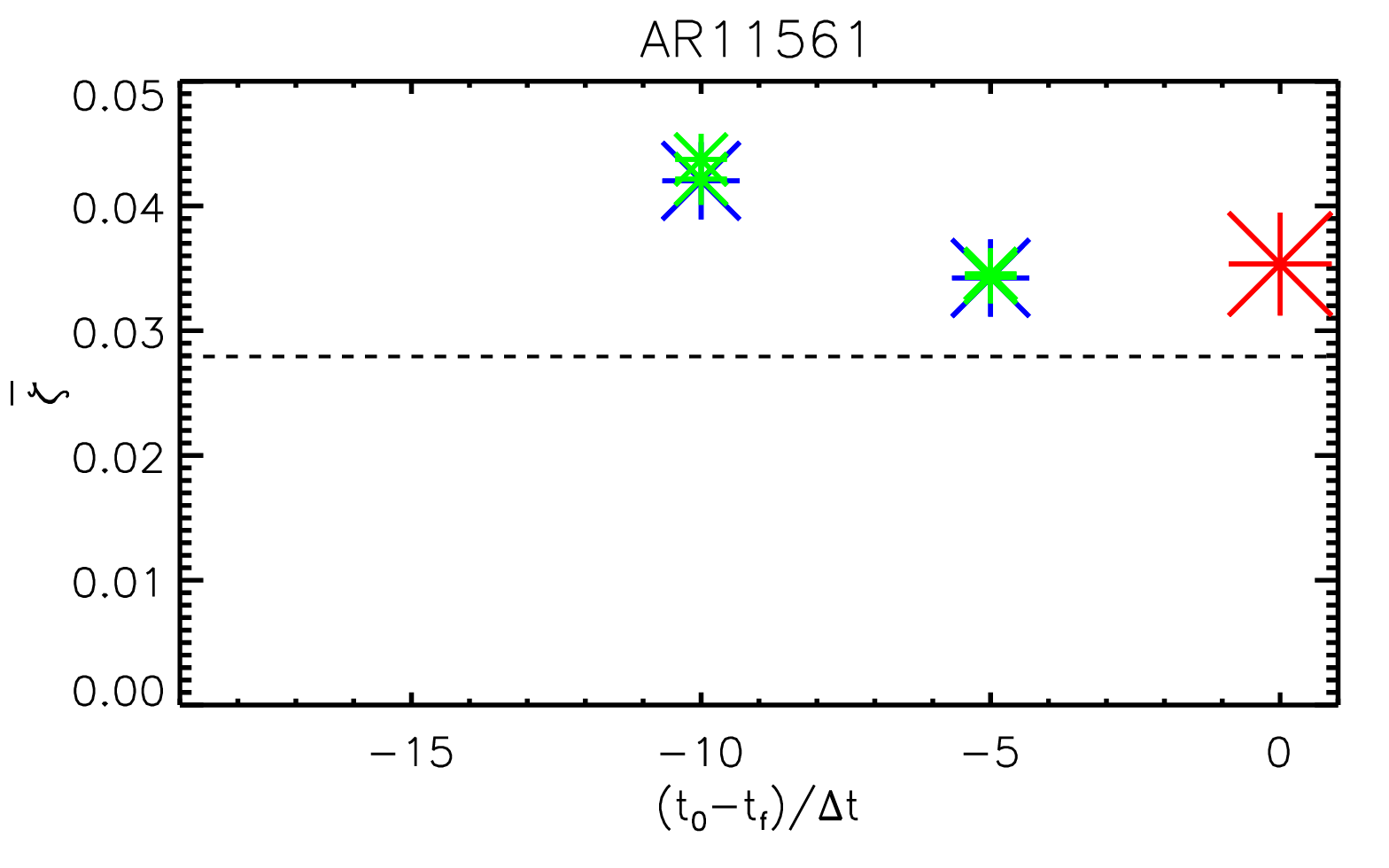}
\includegraphics[scale=0.30]{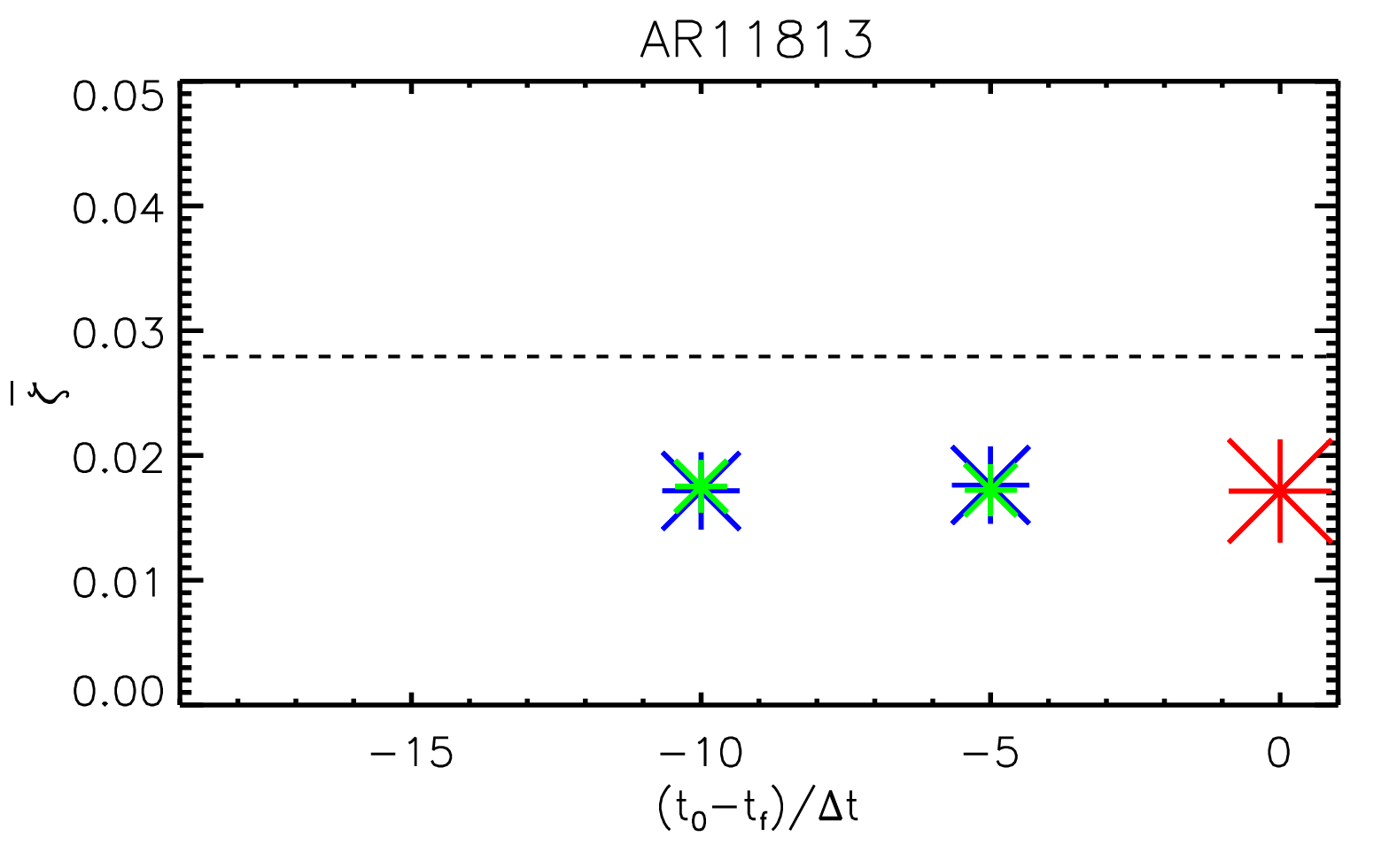}
\caption{The values of $\bar{\zeta}$ for AR11561 (left panel) and AR11813 (right panel) as a function of $t_0$.
The red asterisk represents the simulation where $t_0=t_f$,
the blue asterisks represent the simulations using projected magnetograms with $t_0=t_f-5\Delta t$ and $t_0=t_f-10\Delta t$,
and the green asterisks represent the simulations using the same values of $t_0$ and with added a noise component
of $E^{64}_{noise}=64\sigma_{E}$, $E^{256}_{noise}=256\sigma_{E}$, and $E^{512}_{noise}=512\sigma_{E}$.}
\label{noisecritars}
\end{figure}

Fig.\ref{noisetauevol} shows the evolution of $\zeta_{max}\left(t\right)$ for the two active regions.
The red curve represents the evolution of $\zeta_{max}\left(t\right)$ when $t_0=t_f$
(i.e no noise and only observed magnetograms are used).
The green curves represent the evolution when noise is added.
This is compared to the blue curves which represent
the corresponding unperturbed simulations using projected magnetograms
which deviate from the simulation with $t_0=t_f$ at the same $t_0$.
We find that the simulations with noise do not largely depart from the associated simulations without noise.
The use of projected magnetograms
in the simulations compared with observed magnetograms
has a more prominent effect in departing the evolution of $\zeta_{max}\left(t\right)$.
When we introduce a noise component in the projected magnetogram simulations, there is a marginal effect on the evolution of $\zeta_{max}\left(t\right)$
for AR11561 whereas, there is no significant effect for AR 11813.
Moreover, these minor differences are smoothed out when we focus on the value of $\bar{\zeta}$ (Fig.\ref{noisecritars}),
as we find that the predictions with noise are similar to the associated prediction without noise.
Thus, the distinction between eruptive and non-eruptive active regions is still maintained.
It is clear that the deviations
due to noise are smaller than the variations to the estimation of $\bar{\zeta}$
due to the use of projected magnetograms. This occurs even with
large amplitudes of noise 
with respect to the value of $\sigma_{E}$.
This happens for two main reasons.
On one hand the standard deviation of the electric field from its mean is two orders of magnitude smaller than the electric field acting on the active regions and therefore its small scale fluctuations can only marginally affect the evolution of the magnetic field of the active region.
On the other hand, as the time average of the electric field associated to the noise components tends to zero, it does not significantly affect the values of our eruption metric.
This analysis shows that our projection technique
remains largely insensitive to the introduction of noise and, thus, the underlying mechanisms we are investigating and predicting have a physical nature and are effective in distinguishing eruptive from non-eruptive active regions.

\section{Conclusions}
\label{conclusions}

In this paper we have investigated whether it is possible
to identify eruptive active regions by applying a NLFFF magnetofrictional model \citep{Mackay2011}.
This model uses a sequence of line-of-sight magnetograms to drive the 3D evolution of the 
Sun's magnetic field through a series of quasi-static configurations.
Our study focused on analysing the magnetic field of a sample of active regions to develop an eruption metric,
i.e. a metric that gives the likelihood
that an active region will produce a magnetic flux rope ejection.
We have applied this technique to a set of eight active regions.
Five of these active regions are eruptive,
as observational signatures of an eruption were reported in previous studies,
while the remaining three did not show any observable eruption
and are considered non-eruptive.

In order to distinguish eruptive from non-eruptive active regions, we first derive the quantity $\bar{\zeta}$ from the magnetofrictional simulations
of active region evolution. This quantity is physically linked to: (i) the
presence of magnetic flux ropes, (ii) the intensity and direction of the 
vertical component of the Lorentz force and (iii) the Lorentz force heterogeneity.
We have found that $\bar{\zeta}$ is significantly higher for 
active regions whose evolution shows observational signatures of an eruption.
Therefore, in this work we have empirically identified a threshold that distinguishes the two subsets of active regions.
Future studies must include a larger sample of active regions in order to verify the robustness of this threshold.
When a larger sample of active regions are analysed, it is possible that the two populations of eruptive and non-eruptive regions
overlap. If this occurs then the metric may need to be improved to maintain the discrimination between eruptive and non-eruptive active regions such as we have shown in this work.

For most cases, the distribution of $\zeta\left(x,y,t\right)$ also provides information on the location  of the onset of the eruption.
The key advantage of this work is 
that we use information gained from the full 3D magnetic field configuration of an active region rather than using only the photospheric magnetograms.
We use the full 3D magnetic field configuration,
as we find that the stability of magnetic structures depends 
on the direction and magnitude of the Lorentz force exerted
at the lower photospheric boundary as well as in higher layers of the solar atmoshpere.
However, this model is only effective when the 3D representation
of the active region coronal field is accurate.
Therefore, the evolution of the active region needs to be followed over a long period of time before this approach can be used.

One important aspect of this present study is that the initial magnetic configuration in the magnetofrictional model is assumed to
be potential and this is known to be an oversimplified configuration for active region magnetic fields.
Only the continuous evolution of the photospheric boundary leads to an increasingly non-potential and realistic magnetic field.
To this end, studies have already shown that  
a L5 mission \citep{Gopalswamy2011} can significantly improve our understanding of the solar magnetic field and enable more accurate predictions of flux rope ejections \citep{Mackay2016}
as active regions can be observed over a longer period of time.
Therefore, this study confirms the importance of acquiring a long-lasting data set to reconstruct the coronal field at any given time.
The magnetofrictional model performed significantly better in reproducing the observed global characteristics of the magnetic field configuration when applied to longer data set even when using projected magnetograms.
In the future, we will use NLFFF initial conditions to test
whether we can accurately reproduce the coronal field evolution
using a shorter data set.

To test the robustness of this approach in distinguishing eruptive from non-eruptive active regions we ran
a series of simulations where we varied the time when we stopped
applying observed magnetograms as the lower boundary conditions.
After this time we switched to projected magnetograms that are
derived from the most recent observed magnetograms.
To test the accuracy of the prediction we compared the final value of $\bar{\zeta}$
from these simulations that use projected magnetograms
with the value found in the simulations that
used the full sequence of observed magnetograms.
We find that when the projection is carried out over a time period that is less than $\sim 16$ hours (usually corresponding to 10 magnetograms measurements), 
we reproduce the final value of $\bar{\zeta}$ accurately.
However, the value of $\bar{\zeta}$ of an active region
obtained using the projection technique lies consistently
above or below the empirical threshold 
throughout the majority of the time that  
projected magnetograms are used.
Thus, the identification of eruptive and non-eruptive active regions using
the empirical threshold is robust, even during projection
(having in mind the limitations due to the small sample used).

The results are not significantly affected by the introduction 
of a random noise component.
A number of scientific implications follow from this result.
Firstly, the process that forms solar eruptions in active regions acts
on time scales typically longer than the time interval between two magnetograms used here ($96$ minutes), as the value of $\zeta_{max}\left(t\right)$ follows a continuous evolution and settles to higher value for eruptive active regions.
Secondly, the mechanism that stresses the magnetic field in the build-up to an eruption is also continuous and relatively steady as simulations driven by magnetograms projected
from several magnetograms prior to the eruption time 
show values of our metric $\bar{\zeta}$ largely consistent with the simulation 
driven only by observed magnetograms.
Finally, the mechanism that governs the evolution of active regions
occurs consistently over the entire spatial extent of the active region,
as the introduction of a spatially varying random photospheric noise component
does not lead to a significant alteration of the active region evolution
and in particular to the value of our metric, $\bar{\zeta}$.

In the future, a practical application of the technique described in this paper
will be used to select eruptive active regions as observational targets for
remote sensing instruments on board Solar Orbiter.
This is important, as observational campaigns will need to be planned in advance.
The computational load of this model is light, as
each simulation presented here runs in less than $1$ hour on 16 cores.
This allows the simulations to be run continuously
to monitor the likelihood of an active region erupting
as new magnetograms are acquired.
Moreover, most of the NLFFF simulations presented here are already automated.
Therefore, it is feasible
for the full automation of this approach to be accomplished before the launch of Solar Orbiter or the development of the next generation of space weather models.
For example, we can follow the evolution of a number of active regions on the near side of the Sun and use this approach to select which one Solar Orbiter should observe on the far side during one of its observational campaigns.

In future work, we will explore the possibility of using projected magnetograms
to identify active regions whose likelihood to produce an eruption
is going to increase or decrease in the next few hours.
While this approach is currently not able to identify an exact time when the eruption occurs, by reducing the uncertainty to a few hours
we can still significantly improve our Space Weather forecast tools.
Operational Space Weather development is a very dynamic research area
where several tools have been recently developed. While our results are interesting 
it remains to be seen whether any future operational tool based on
this technique can perform better than the existing ones.
For example, established techniques to predict solar flares are becoming increasingly more common in operations.
Two of these tools include MAG4 \citep{2011SpWea...9.4003F,2014AAS...22440204F} which is used by the National Oceanic and Atmospheric Administration (NOAA) and by NASA's Space Radiation Analysis Group (SRAG) and the FLARECAST \citep{2017AGUFMSA21C..07G,2018cosp...42E1181G}
platform that is developed by a consortium of Europe based institutions and includes 
some follow up applications using machine learning \citep{2018SoPh..293...28F}.

There are some steps that we can undertake
in the future to improve this approach.
Firstly, a more sophisticated projection technique will improve 
our capability to simulate the evolution of active regions.
Secondly, it is possible that a wider study involving a larger sample of active regions (both erupting and non-erupting)
will place more stringent conditions on values of $\bar{\zeta}$.
Certainly, the long term effect of flux emergence on $\bar{\zeta}$
needs to be mitigated.
Additionally, more physical conditions can be 
included such as the torus instability criteria
\citep{TorokKliem2005, Aulanier2010,Zuccarello2015},
where the decay of the field overlying the flux rope with height is calculated using the decay index.
Also, this approach would benefit from the automatic detection of magnetic flux
ropes such as that used in \citet{Lowder2017}.
These improvements are likely to make the technique more robust and effective.

\acknowledgments

This research
has received funding from the Science and Technology Facilities Council (UK)
through the consolidated grant ST/N000609/1 and the European Research Council
(ERC) under the European Union Horizon 2020 research and innovation
program (grant agreement No. 647214).
This work used the DiRAC@Durham facility managed by the Institute for
Computational Cosmology on behalf of the STFC DiRAC HPC Facility
(www.dirac.ac.uk). The equipment was funded by BEIS capital funding
via STFC capital grants ST/P002293/1, ST/R002371/1 and ST/S002502/1,
Durham University and STFC operations grant ST/R000832/1. DiRAC is
part of the National e-Infrastructure.
S.L.Y. would like to acknowledge STFC for support via the Consolidated Grant SMC1/YST025 and SMC1/YST037.
DHM would like to thank both the UK STFC and the ERC for financial support.

%



\bibliographystyle{aasjournal}
\bibliography{ref}


\end{document}